\documentclass[twocolumn]{aa}
\usepackage{graphicx}
\usepackage{txfonts}
\usepackage{rotating}
\usepackage{aalongtable,lscape}
\def\XMM{{\sl XMM-Newton}}

\def\Chan{{\sl Chandra}}
\def\cgs{erg cm$^{-2}$ s$^{-1}$}
\def\SXMM{{\tt 2XMM}}

\begin{document}
   \title{High-precision multi-band measurements of the angular clustering of X-ray sources}

   \author{J. Ebrero\inst{1,2}
          \and
          S. Mateos\inst{1}
          \and
          G. C. Stewart\inst{1}
          \and
          F. J. Carrera\inst{2}
          \and
          M. G. Watson\inst{1}
          }

   \offprints{J. Ebrero, \email{ecarrero@ifca.unican.es, jec33@star.le.ac.uk}}

   \institute{Department of Physics and Astronomy, University of Leicester, University Road, LE1 7RH, Leicester, UK
     \and
     Instituto de F\'\i sica de Cantabria (CSIC-UC), Avenida de los
     Castros, 39005 Santander, Spain     
             }

   \date{Received <date> / Accepted <date>}

   \abstract
% Context
 {}
% Aims
 {We aim to study the large-scale structure of an extragalactic serendipitous X-ray survey with unprecedented accuracy thanks to the large statistics involved, and provide insight into the environment of AGN at the epochs when their space density declines ($z \sim 1-2$).}
% Methods
 {In this paper we present the two-point angular correlation function of the X-ray source population of 1063 \XMM{} observations at high Galactic latitudes, comprising up to $\sim$30000 sources over a sky area of $\sim$125.5~deg$^{2}$, in three energy bands: 0.5-2 (soft), 2-10 (hard), and 4.5-10 (ultrahard)~keV. This is the largest survey of serendipitous X-ray sources ever used for clustering analysis.}
% Results
 {We have measured the angular clustering of our survey and find significant positive clustering signals in the soft and hard bands ($\sim$10$\sigma$ and $\sim$5$\sigma$, respectively), and a marginal clustering detection in the ultrahard band ($<$1$\sigma$). We find dependency of the clustering strength on the flux limit and no significant differences in the clustering properties between sources with high hardness ratios (and therefore likely to be obscured AGN) and those with low hardness ratios. We deprojected the angular clustering parameters via Limber's equation to compute their typical spatial lengths. From that we have inferred that AGN at redshifts of $\sim$1 are embedded in dark matter haloes with typical masses of $\langle \log M_{DMH}\rangle \simeq 12.60\pm0.34$~$h^{-1}$~M$_\odot$ and lifetimes of $t_{AGN}=3.1-4.5 \times 10^8$~yr.}
% Conclusions
 {Our results show that obscured and unobscured objects share similar clustering properties and therefore they both reside in similar environments, in agreement with the unified model of AGN. The short AGN lifetimes derived suggest that AGN activity might be a transient phase that can be experienced several times by a large fraction of galaxies throughout their lives.}


   \authorrunning{J. Ebrero et al.}
   \titlerunning{High-precision multi-band measurements of the angular clustering of X-ray sources}
   
   \maketitle
%
%________________________________________________________________

\section{Introduction}

Active galactic nuclei (AGN) are the brightest persistent extragalactic sources known, with their X-ray emission the most common feature among them. Thanks to their large bolometric output, AGN can be detected through cosmological distances, which makes them essential tracers of galaxy formation and evolution, as well as the large-scale structure of the Universe. Clustering studies of AGN at redshift $\sim$1, when strong structure formation processes took place, are key tools for understanding the underlying mass distribution and evolution of cosmic structures (Hartwick \& Schade \cite{Hartwick90}).

Moreover, a number of works suggest that the most luminous AGN were formed earlier in the Universe, while the less luminous ones were formed later (Ueda et al. \cite{Ueda03}, Hasinger et al. \cite{Hasinger05}, Barger et al. \cite{Barger05}, La Franca et al. \cite{LaFranca05}, Silverman et al. \cite{Silverman08}, Ebrero et al. \cite{Ebrero08}). Understanding the large-scale clustering of AGN will therefore provide more clues to the environment of AGN at these epochs and to how it is linked to the formation and accretion of matter onto the central supermassive black hole, since it is thought to be triggered by major mergers or close interactions between galaxies (i.e. Kauffmann \& Haehnelt \cite{Kauffmann00}, Cavaliere \& Vitorini \cite{Cavaliere02}, Menci et al. \cite{Menci04}, Di Matteo et al. \cite{DiMatteo05}, Granato et al. \cite{Granato06}).

The simplest way to measure clustering is the two-point angular correlation function, which measures the excess probability of finding a pair of sources at a given angular distance. There have been multiple determinations of the angular correlation function of both optically and X-ray selected AGN (Vikhlinin \& Forman \cite{Vikhlinin95}, Akylas et al. \cite{Akylas00}, Giacconi et al. \cite{Giacconi01}, Basilakos et al. \cite{Basilakos04}, Basilakos et al. \cite{Basilakos05}, Gandhi et al. \cite{Gandhi06}, Carrera et al. \cite{Carrera07}, Miyaji et al. \cite{Miyaji07}, Ueda et al. \cite{Ueda08}). These works have shown that AGN detected in soft X-rays (0.5-2~keV) tend to cluster strongly. Nevertheless, at higher energies many of these results were inconclusive because of the small-number statistics, ranging from marginally significant to no clustering detections at all.

The angular correlation function, however, only measures overdensities projected in the sky thus blurring the underlying spatial structure. With larger and more accurate spectroscopic identification campaigns of AGN, an increasing number of works have calculated the spatial clustering of these sources (e.g. Mullis et al. \cite{Mullis04}, Gilli et al. \cite{Gilli05}, Yang et al. \cite{Yang06}, Gilli et al. \cite{Gilli09}). The clustering properties derived from these determinations have been used to evaluate the evolution of AGN clustering with redshift, and the mass of dark matter haloes (DMH) in the context of a cold dark matter (CDM) scenario, in which DMH of different mass cluster differently.

To accurately measure the angular correlation function, a sample that achieves both width (to prevent biases caused by a single structure) and depth (to ensure high angular density) is needed. Moreover, X-ray selected samples are less biased against obscuration than optically selected samples, especially in the 2-10~keV energy range, and are therefore an ideal resource for large-scale structure and evolutionary analysis. In this paper we use 1063 \XMM{} observations at high Galactic latitudes over a sky area of 125~deg$^2$ (Mateos et al. \cite{Mateos08}) to compute the angular correlation function of serendipitous X-ray sources in three energy bands: 0.5-2~keV, 2-10~keV and 4.5-10~keV. The sample comprises over $\sim$30,000 sources, thus being the largest compiled sample to investigate clustering so far.

This paper is organised as follows: in section~\ref{sample} we overview the sample used in this work and describe its general properties. In section~\ref{cosmicvariance} we perform a crude analysis similar to the traditional counts-in-cells methods as a preliminary test for clustering. In section~\ref{angular} we undertake the calculation of the angular correlation function of our sample in different energy bands, while in section~\ref{spatial} we study the deprojection to the real space of our results via the inversion of Limber's equation. Finally, our conclusions are reported in section~\ref{conclusions}.

Throughout this paper we have assumed a cosmological framework with $H_0=70$~km s$^{-1}$ Mpc$^{-1}$, $\Omega_M=0.3$ and $\Omega_\Lambda=0.7$ (Spergel at al. \cite{Spergel03}). 

%__________________________________________________________________
\section{The X-ray data}
\label{sample}

%-----------------------------------------
% Sky area figures
\begin{figure}
\centering
\hbox{
\includegraphics[width=6.5cm,angle=270.0]{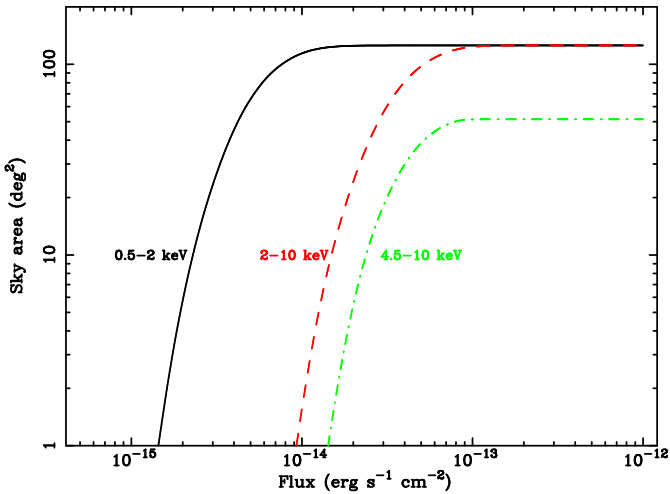}
}
\caption[]{Sky area of the survey as a function of flux in the soft ({\it solid line}), hard ({\it dashed line}) and ultrahard ({\it dot-dashed line}) bands.}
\label{fig:skyareas}
\end{figure}
%-------------------------------------------

In this work we use the sample presented in Mateos et al. (\cite{Mateos08}). The observations are a subset of those employed in producing the second \XMM{} serendipitous source catalogue, \SXMM{}\footnote{{\tt http://xmmssc-www.star.le.ac.uk/Catalogue/2XMM/}} (Watson et al. \cite{Watson08}). \SXMM{} is composed by observations from the European Photo Imaging Cameras (EPIC) on board \XMM{}, although this sample was built using data from the EPIC-pn camera only (Turner et al. \cite{Turner01}). 
%All the observations were reprocessed using the same pipeline configuration to guarantee an uniform data set, filtering high particle background periods by excluding the time intervals in which the 7-15~keV pn count rate was higher than 10 cts/arcmin$^2$/ks (see Mateos et al. \cite{Mateos08} for details).

The selected observations fulfill the following criteria:

\begin{enumerate}
  \item High galactic latitude fields ($\left|b\right|>20^\circ$) in order to minimize the contamination from Galactic sources.
  \item Fields with at least 5~ks of clean exposure time.    
  \item Fields free of bright and/or extended X-ray sources. 
\end{enumerate}

If there were observations carried out at the same sky position, the overlapping area from the observation with the shortest clean exposure time was removed. The resulting sample comprised 1129 observations. For the purposes of this work we have also removed the observations belonging to the Virgo Cluster, M31, M33, Large Magellanic Cloud and Small Magellanic Cloud fields, ending up with a final sample of 1063 observations.

The analysis presented in this paper has been carried out in the following energy bands:

\begin{itemize}
\item {\it Soft}: 0.5-2~keV

\item {\it Hard}: 2-10~keV.

\item {\it Ultrahard}: 4.5-10~keV
\end{itemize}

Sources detected in each energy band have a minimum detection likelihood of 15 (which roughly corresponds to a 5$\sigma$ significance of detection, Cash \cite{Cash79}) and fluxes $<10^{-12}$~\cgs{}. The targets of the observations were removed.

The sky coverage as a function of the X-ray flux was calculated using an empiral approach, computing a sensitivity map for each observation that provides the minimum count rate required for a source to be detected at each position in the field of view, taking into account the local effective exposure and the background level. The procedure is described in detail in Carrera et al. (\cite{Carrera07}) and Mateos et al. (\cite{Mateos08}). Sources with actual count rates below the sensitivity map value at their position were therefore excluded from the analysis for consistency with the sky area calculations. The fraction of sources removed this way is less than 4\% in the 0.5-2~keV band, $\sim$5\% in the 2-10~keV band, and $\sim$7\% in the 4.5-10~keV band. Similarly as in Mateos et al. (\cite{Mateos08}), we have also only considered sources that were detectable over a minimum area of $\Omega_{min}=1$~deg$^2$ in order to avoid uncertainties due to the low count statistics, or inaccuracy in the sky coverage calculation at the very faint detection limits. Less than 0.5\% of the sources in the soft and hard bands were removed this way, while in the ultrahard band this fraction raises to $\sim$1.5\%. Changes in the flux limits of the survey were negligible.

Since the density of sources per field is a key issue to the study of angular clustering (a low source density can lead to a no-clustering signal even if there is an actual clustering present, see Carrera et al. \cite{Carrera07}), we have removed the fields with exposure times below 20~ks from the analysis in the ultrahard band. This way we enhance the source density in this band from $\sim$1 to $\sim$4 sources per field (still far from the source density in the soft and hard bands, $\sim$30 and $\sim$10 sources per field, respectively), for a total of 432 fields.

Hence, the overall sky coverage of the sample is 125.52~deg$^2$ comprising 31288 and 9188 sources in the softand hard bands, respectively. The sky area covered by the $>$20~ks ultrahard sample is $\sim$51.5~deg$^2$ for a total of 1259 sources (see Figure~\ref{fig:skyareas}). The properties of the sample are summarised in Table~\ref{tab:sample}.

%-------------------------
% Table of samples
\begin{table}
  \centering
  \caption[]{Sample summary.}
  \label{tab:sample}

  \begin{tabular}{lcccc}
    \hline
    \hline
    \noalign{\smallskip}
    Band (keV) &  $N_{obs}$$^a$ & $N_{sou}$$^b$ & Flux limit   & Area       \\
               &                &               &  (\cgs{})    & (deg$^{2}$) \\
    \noalign{\smallskip}
    \hline
    \noalign{\smallskip}
    Soft (0.5-2)       &  1063 & 31288 & 1.4$\times$10$^{-15}$  & 125.52  \\
    Hard (2-10)        &  1063 & 9188  & 9.2$\times$10$^{-15}$  & 125.52   \\
    Ultrahard (4.5-10)$^c$ &  432  & 1259  & 1.4$\times$10$^{-14}$  & 51.47   \\
    \noalign{\smallskip}
    \hline
    \noalign{\smallskip}
    \multicolumn{4}{l}{$^a$ Number of \XMM{} observations used}\\
    \multicolumn{4}{l}{$^b$ Number X-ray sources used}\\
    \multicolumn{5}{l}{$^c$ For observations with exposure times $>$20~ks only}\\
  \end{tabular}
\end{table}
%------------------------------------------

%-----------------------------------------
% LogNlogS
\begin{figure}
\centering
\hbox{
\includegraphics[width=6.5cm,angle=270.0]{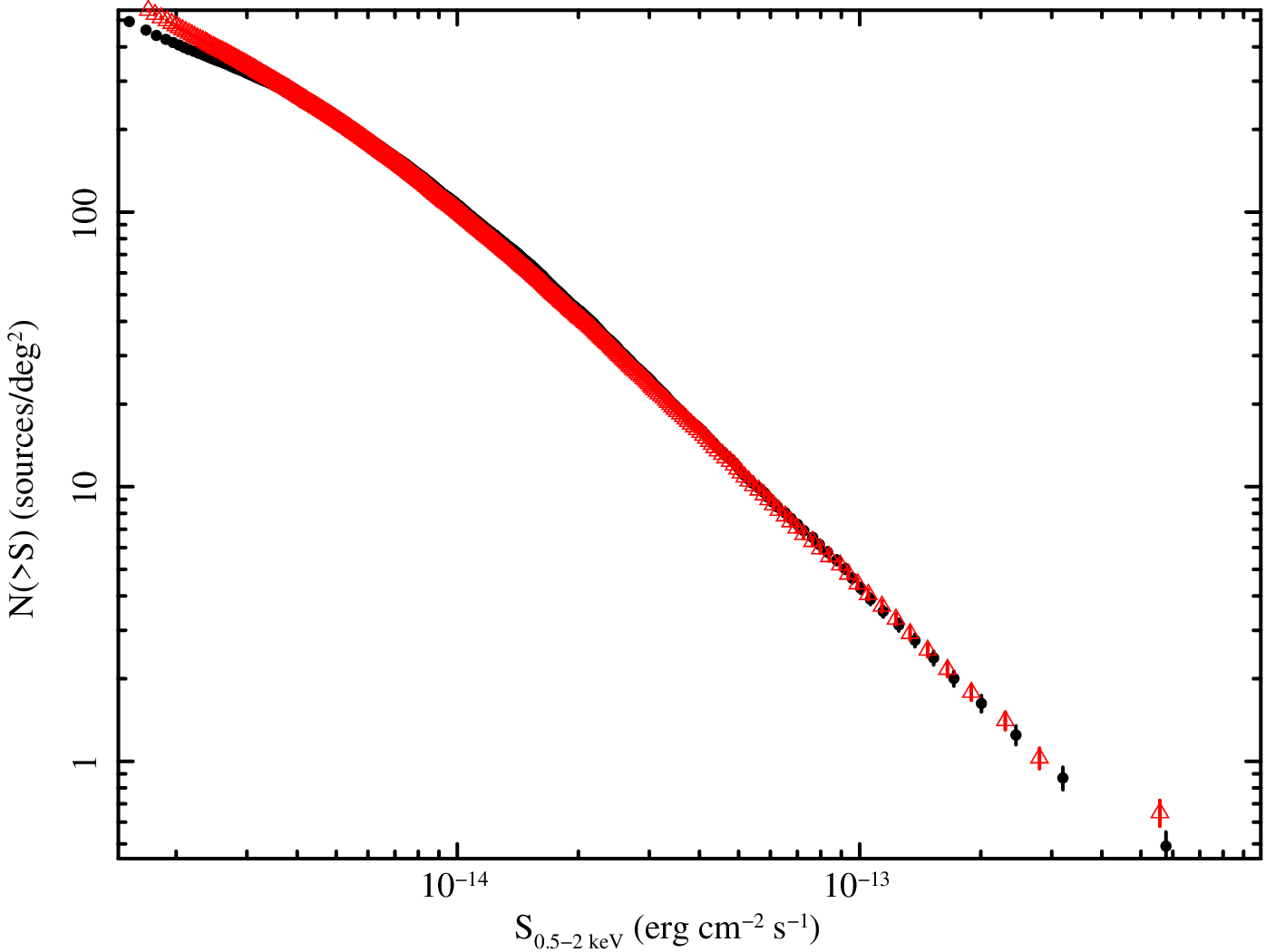}
}
\hbox{
\includegraphics[width=6.5cm,angle=270.0]{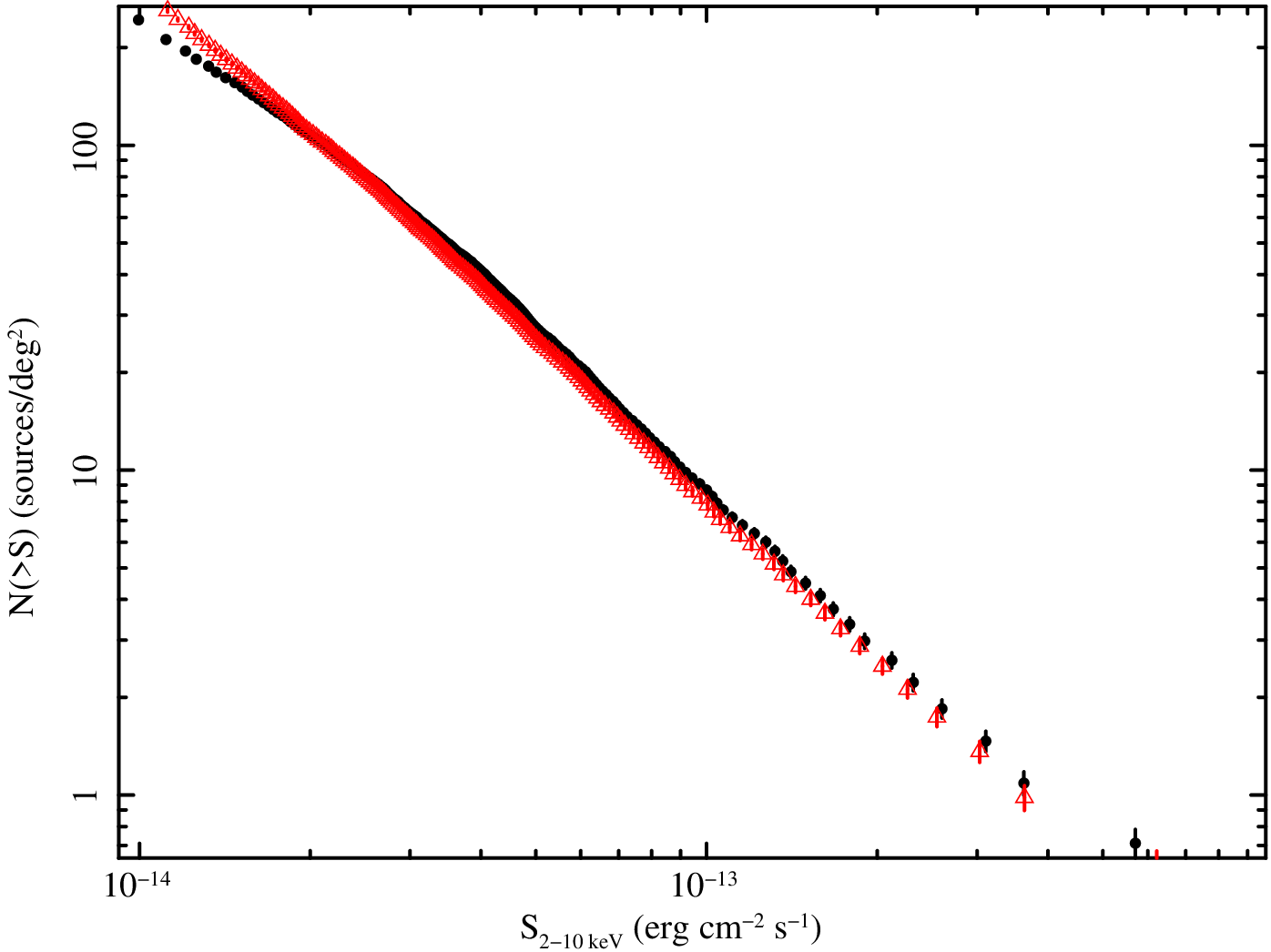}
}
\caption[]{Source counts of the real ({\it triangles}) and simulated samples ({\it dots}) in the soft ({\it upper panel}) and hard ({\it lower panel}) bands.}
\label{fig:lognlogs}
\end{figure}
%-------------------------------------------

%_________________________________________________

\section{Cosmic variance}
\label{cosmicvariance}

As a preliminary test for source clustering, we have compared the actual number of sources detected in each field $N_k$ with the number of expected sources $\lambda_k$ obtained from the best fit $\log N - \log S$ of Mateos et al. (\cite{Mateos08}). This is similar to the traditional count-in-cells method. If there is a cosmic structure present in a few (or most of the) fields, the number of sources detected would be significantly different compared to that of a random uniform distribution as measured by the overall $\log N - \log S$.

For this purpose we have used the cumulative Poisson distributions

\begin{equation}
\begin{array}{lcll}
P_{\lambda_k}(\geq N_k) & = & \sum_{l=N_k}^{\infty}P_{\lambda_k}(l) & ,N_k>\lambda_k \nonumber\\
P_{\lambda_k}(\leq N_k) & = & \sum_{l=0}^{N_k}     P_{\lambda_k}(l) & ,N_k<\lambda_k, \\
\end{array}
\label{eq:cumpois}
\end{equation}

\noindent where $P_{\lambda_k}(l)$ is the Poisson probability of detecting $l$ sources when the expected number of sources is $\lambda$. The likelihood statistics for the whole sample is hence

\begin{equation}
L'=-2\sum_{k}\ln P_{\lambda_k}(\geq N_k) - 2\sum_{k'}\ln P_{\lambda_{k'}}(\leq N_{k'}),
\label{eq:lik}
\end{equation}

\noindent where $k$ runs over the fields for which $N_k > \lambda_k$ and $k'$ runs over the fields for which $N_k' < \lambda_k'$. This procedure is the same as the one used in Carrera et al. (\cite{Carrera07}), and differs slightly from that of Carrera et al. (\cite{Carrera98}) which uses $P_{\lambda}(l)$ instead of the cumulative probabilities.

The observed $L'$ values thus obtained were compared to 1000000 simulated likelihood values calculated using $\lambda_k$ and Poisson statistics, finding that the number of simulations with likelihood values above the observed ones were 0, 6522 and 703713 for the soft, hard and ultrahard bands, respectively. It can be seen that there are significant deviations from the expected number of sources from a random uniform distribution in the soft and hard bands.

The fact that all the simulations in the soft band show better likelihoods than our sample indicates that we can set up a lower limit for the significance of the deviation (i.e. an evidence for clustering) of at least $\gtrsim$6$\sigma$, whereas in the hard band the deviation is of the order of $\sim$3$\sigma$. The small number of simulations in the ultrahard band with likelihoods better than the observed ones suggests that no significant deviations (well below 1$\sigma$) from the random distribution were found, probably due to the large statistical noise (Stewart et al., in preparation). However, these results point out in the direction that a cosmic large-scale structure might be present in most of the observations, therefore setting ground to more ellaborate clustering analysis.

%-----------------------------------------
% wtheta soft, hard, uh
\begin{figure*}
\centering
\hbox{
\includegraphics[width=6.5cm,angle=270.0]{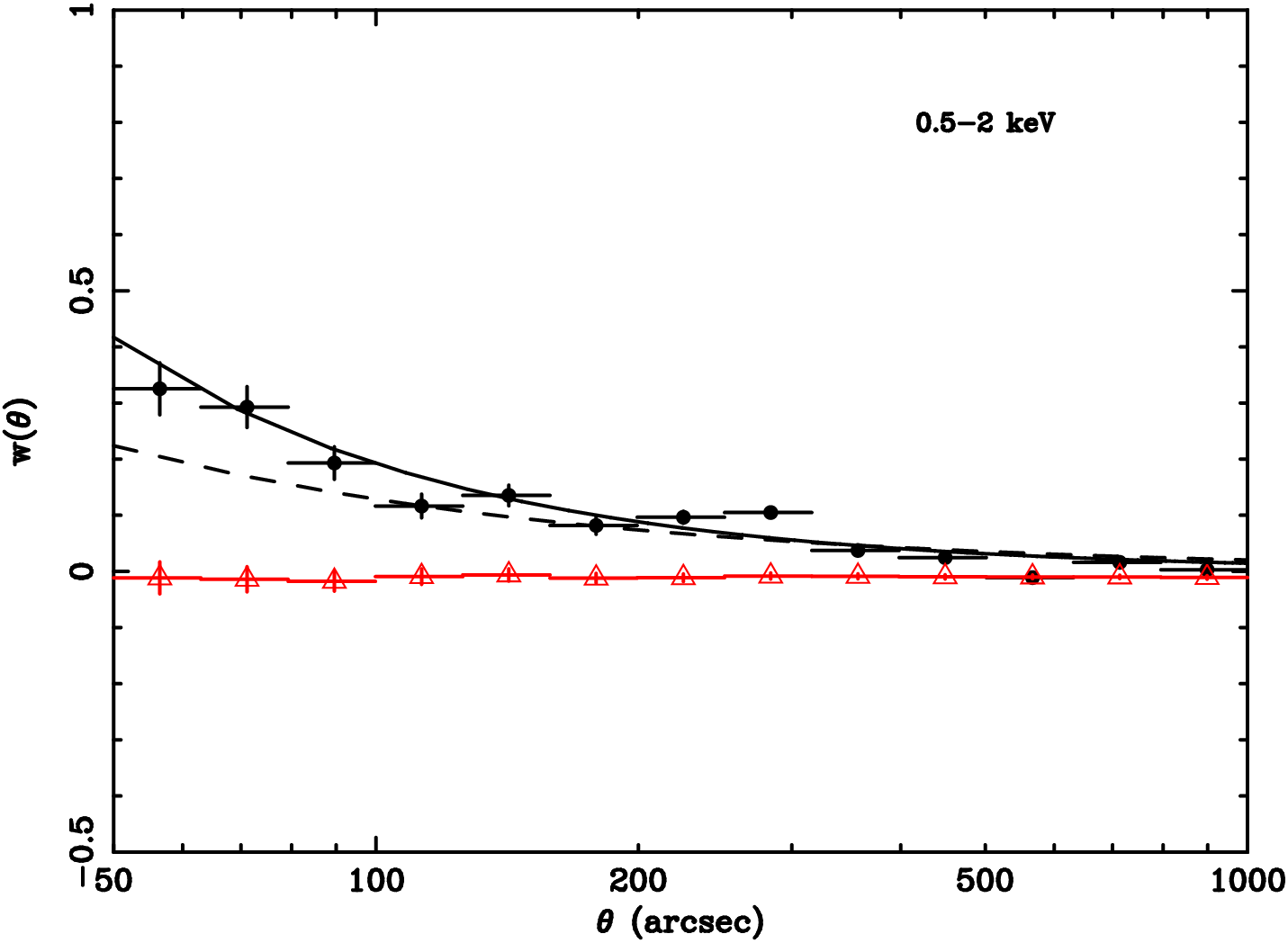}
\includegraphics[width=6.5cm,angle=270.0]{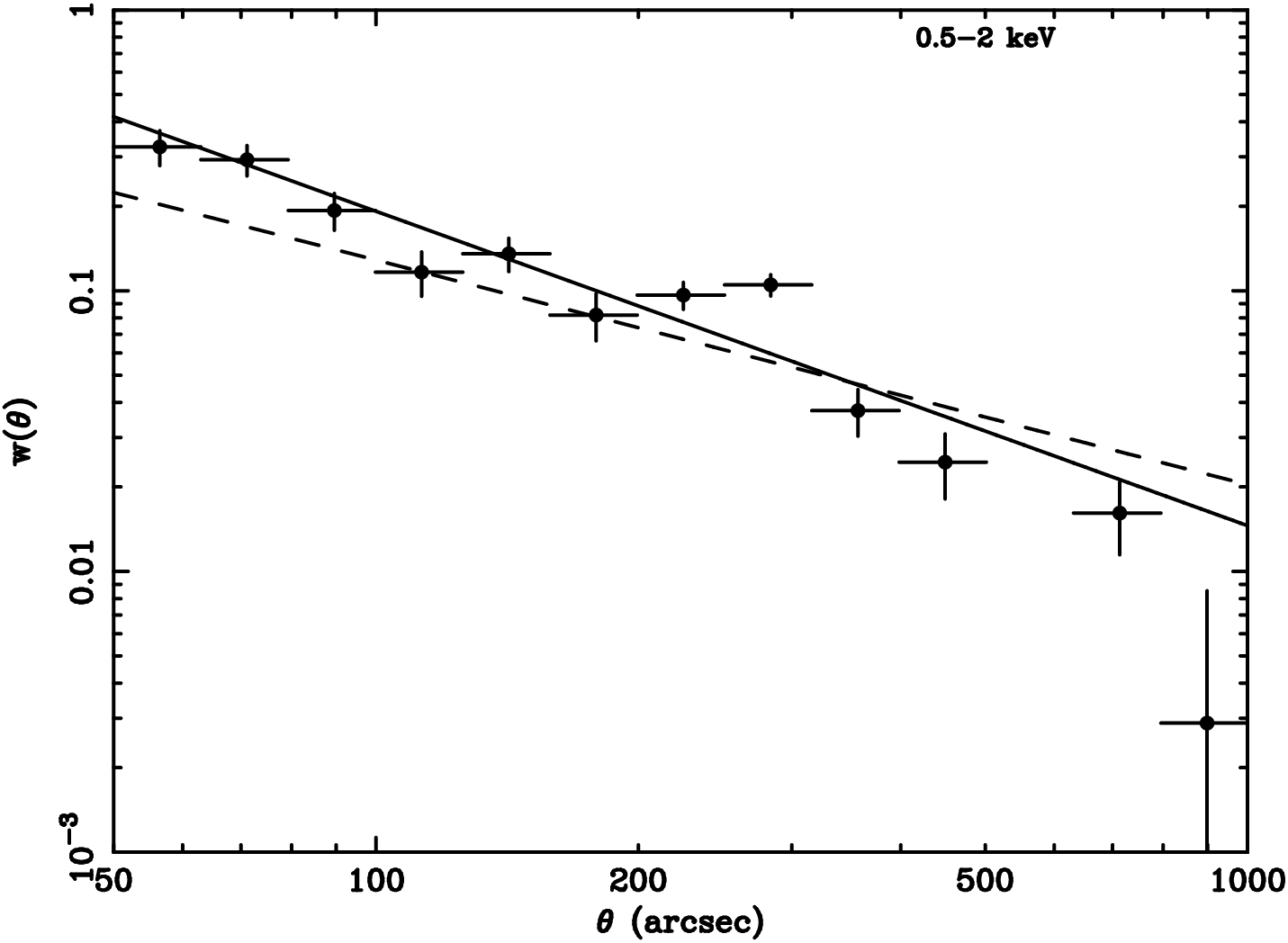}
}
\hbox{
\includegraphics[width=6.5cm,angle=270.0]{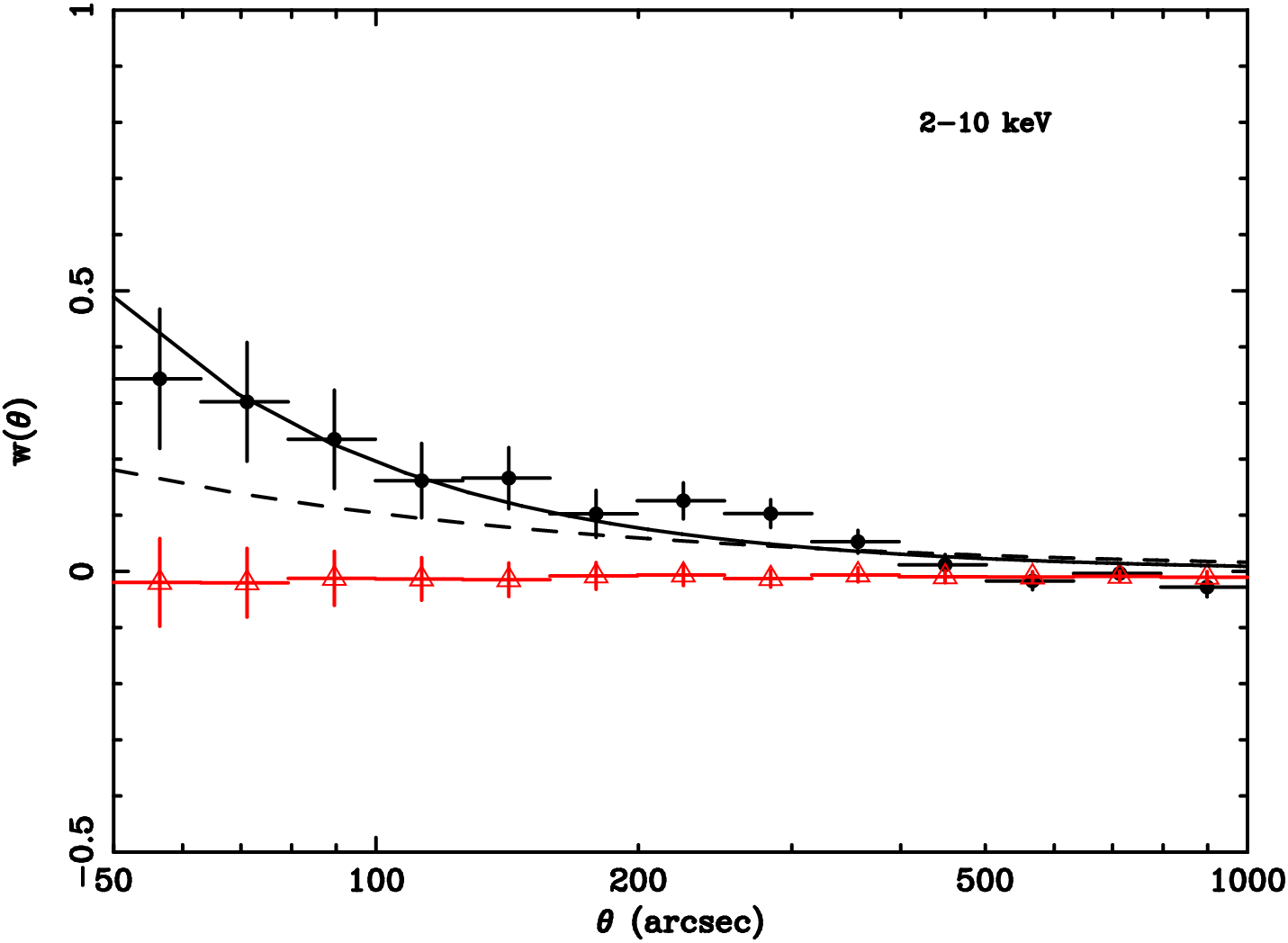}
\includegraphics[width=6.5cm,angle=270.0]{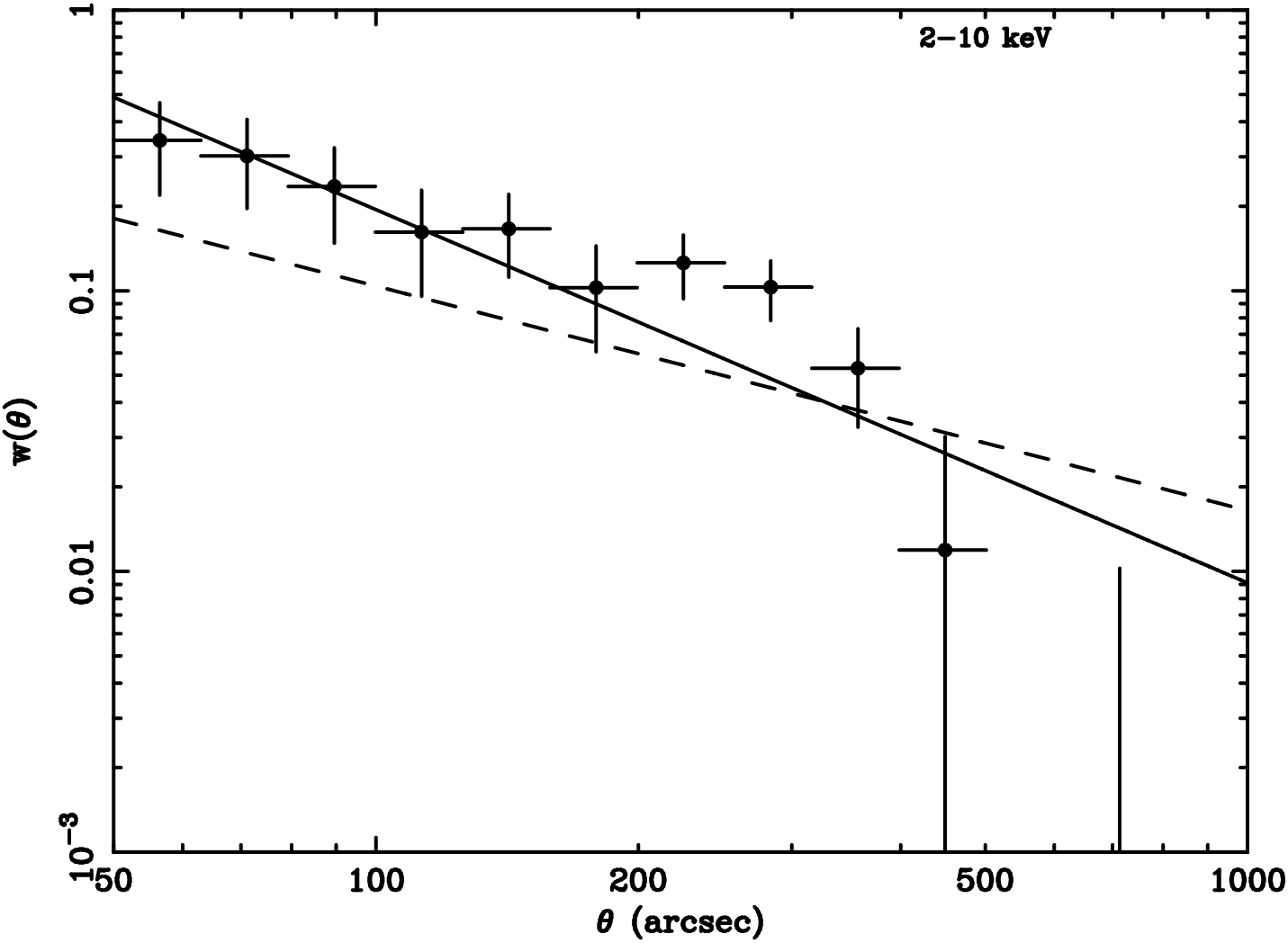}
}
\hbox{
\includegraphics[width=6.5cm,angle=270.0]{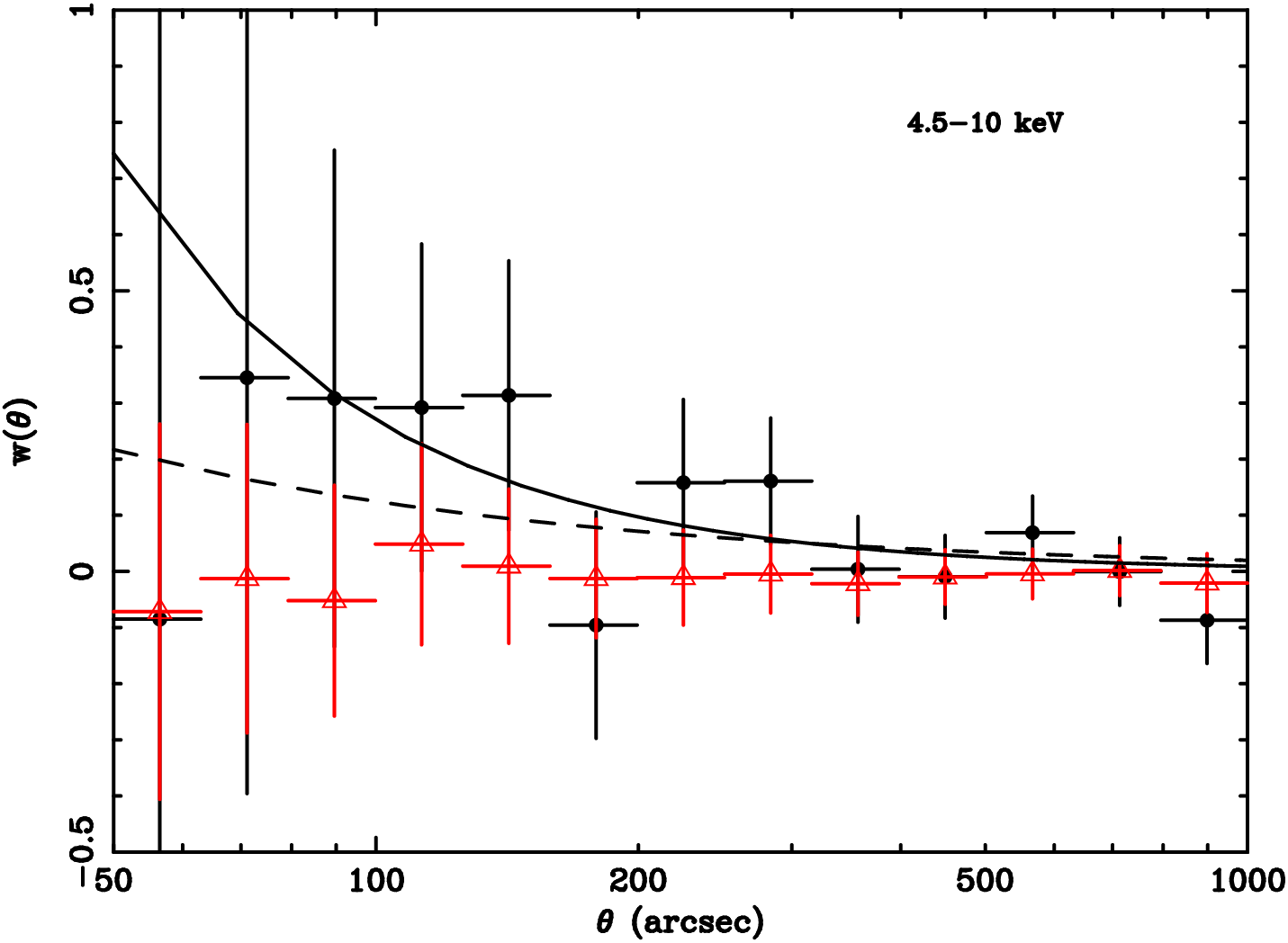}
\includegraphics[width=6.5cm,angle=270.0]{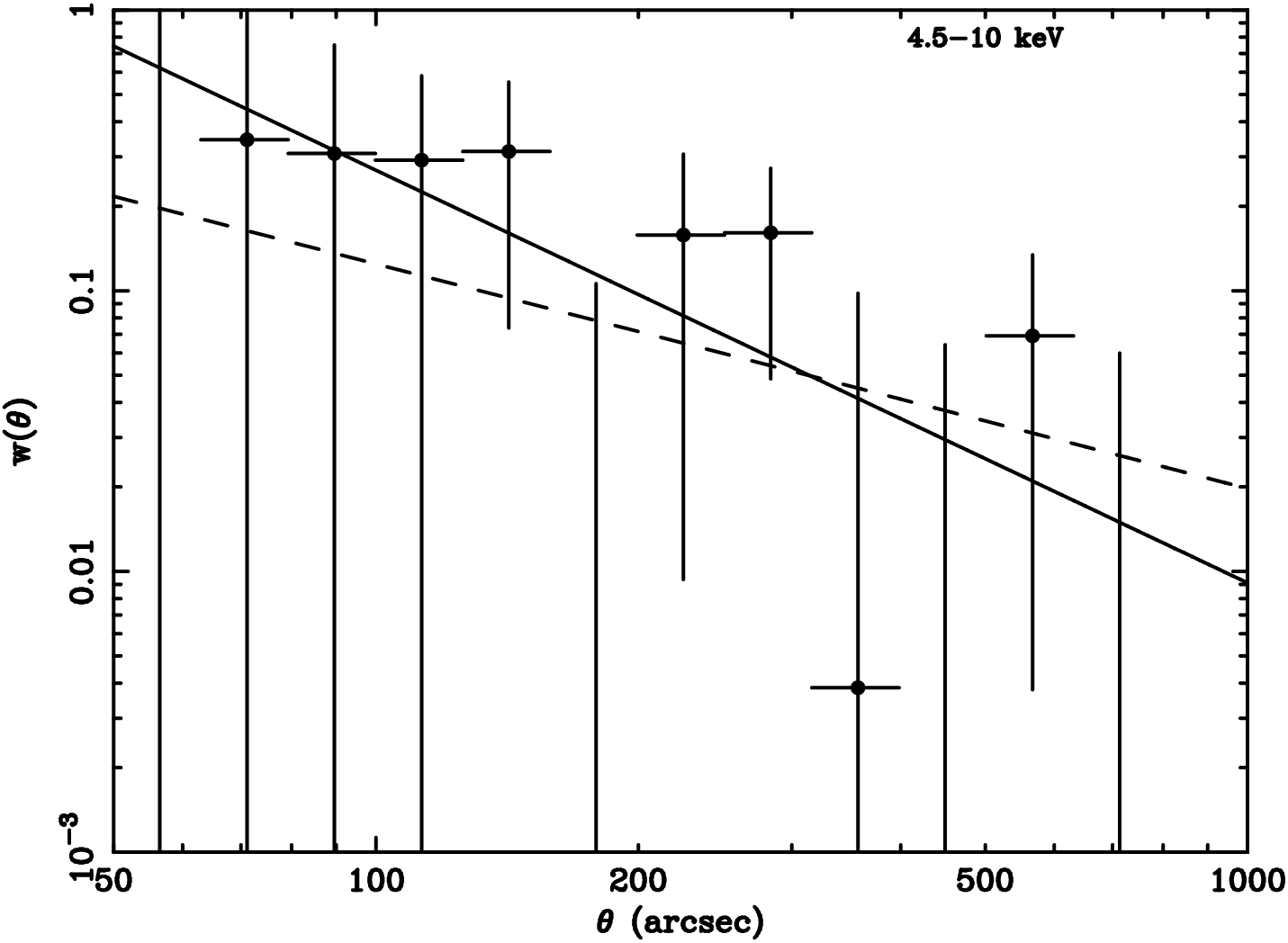}
}
\caption[]{Angular correlation function in the soft ({\it Top panels}), hard ({\it Centre panels}) and ultrahard ({\it Bottom panels}) bands. Left hand panels are represented in linear-log scales, while right hand panels are represented in log-log scales. Solid dots are the observed data. Solid triangles represent the average random estimation of $w(\theta)$ used to calculate the integral constraint (see text). Overplotted is the best-fit $\chi^2$ with and without fixed slope (dashed and solid lines, respectively).}
\label{fig:wtheta}
\end{figure*}
%-------------------------------------------

%_________________________________________________

\section{The angular correlation function}
\label{angular}

The two-point angular correlation function $w(\theta)$ determines the joint probability of finding two objects in two small angular regions $\delta\Omega_1$ and $\delta\Omega_2$ separated by an angular distance $\theta$ with respect to that of a random distribution (Peebles \cite{Peebles80}),

\begin{equation}
\delta P = n^2\delta\Omega_1\delta\Omega_2[1+w(\theta)],
\label{eq:deltaP}
\end{equation}

\noindent where $n$ is the mean sky density of objects. If $w(\theta)=0$ the distribution is homogeneous.

The angular separation is a projection in the sky of the actual spatial separation between two sources at different redshifts thus somewhat blurring the true underlying spatial clustering, which needs accurate and highly complete redshift determinations to be properly measured. The angular correlation function is, nevertheless, a powerful approach given the large size of the present two-dimensional extragalactic sample.

The spatial clustering, however, can be estimated by deprojecting the computed angular correlation function assuming a given redshift distribution (which can be either empirically measured or derived from a luminosity function model) via the inversion of Limber's equation (see section~\ref{limber}).

\subsection{Method}
\label{method}

To calculate the angular correlation function we have used the estimator proposed by Landy \& Szalay (\cite{Landy93}):

\begin{equation}
\label{eq:estim}
w(\theta_i)=\frac{DD-2DR+RR}{RR},
\end{equation}

\noindent where $DD$, $DR$ and $RR$ are the normalised number of pairs of sources in the $i$-th angular bin for the Data-Data, Data-Random and Random-Random samples, respectively. $DD$, $DR$ and $RR$ are normalised dividing them by the total number of pairs in the sample:

\begin{equation}
\label{eq:norm}
\begin{array}{l}
f_{DD}=N_D(N_D-1)/2 \\
f_{DR}=N_DN_R \\
f_{RR}=N_R(N_R-1)/2. \\
\end{array}
\end{equation}

To produce the random source sample against which we have compared the real source sample searching for overdensities at different angular distances, we have tried to mimic as closely as possible the real distribution of the detection sensitivity of the survey. The method employed here is similar to that used in Carrera et al. (\cite{Carrera07}) and can be summarised as follows:

\begin{enumerate}
  \item We formed a pool that included all the real sources with detection likelihoods $\geq$15 in the band under study, irrespective of whether their count rates were above the sensitivity map of their corresponding field at the source position or not.
  \item For each field, we extracted sources at random from this pool keeping their original count rates and distances to the optical axis of the X-ray telescope but randomizing their azimuthal angle around it.
  \item If the source had a count rate above the sensitivity map of the field under consideration at the new position, it was kept in the random sample. Otherwise, the source was discarded and a new source was drawn from the pool until the number of valid simulated sources matched the number of real detected sources in the field.
\end{enumerate}

%-------------------------
% Table of fits
\begin{table}
  \centering
  \caption[]{Summary of the fits.}
  \label{tab:fits}

  \begin{tabular}{lcccc}
    \hline
    \hline
    \noalign{\smallskip}
    Band (keV) &  $N_{sou}$ & $\theta_0$   & $\gamma-1$ & $\chi^2$/d.o.f.    \\
               &            &  (arcsec)  &   \\
    \noalign{\smallskip}
    \hline
    \noalign{\smallskip}
    Soft (0.5-2)   &   31288 & 22.9$\pm$2.0 & 1.12$\pm$0.04 & 8.4  \\
%       &   &   & 26.0$\pm$2.0 & 1.29$\pm$0.04  &  No  \\
       &    & 7.7$\pm$0.1 & 0.8 $(fixed)$ & 12.3  \\
%       &   &   & 5.3$\pm$0.1 & 0.8 $(fixed)$  &  No  \\
    \hline
    \noalign{\smallskip}
    Hard (2-10)  &  9188  & 29.2$_{-5.7}^{+5.1}$ & 1.33$_{-0.11}^{+0.10}$ & 1.9   \\
%      &  &   & 30.7$_{-5.7}^{+5.1}$ & 1.51$_{-0.13}^{+0.12}$  & No   \\
      &    & 5.9$\pm$0.3 & 0.8 $(fixed)$ & 3.0  \\
%      &  &   & 3.9$\pm$0.2 & 0.8 $(fixed)$  & No   \\
    \hline
    \noalign{\smallskip}
    Ultrahard (4.5-10) &  1259  & 40.9$_{-29.3}^{+19.6}$ & 1.47$_{-0.57}^{+0.43}$ & 0.5   \\
%     &  &   & 42.3$_{-30.1}^{+18.7}$ & 1.64$_{-0.64}^{+0.49}$  &  No   \\
     &    & 7.4$\pm$1.4 & 0.8 $(fixed)$ & 0.6  \\
%     &  &   & 5.5$\pm$1.3 & 0.8 $(fixed)$  &  No   \\
    \noalign{\smallskip}
    \hline
  \end{tabular}
\end{table}
%------------------------------------------

This method allowed us to reproduce the decline of the detection sensitivity with the off-axis angle in the simulated sample. We have performed 100 simulations of the whole sample in each band this way. For comparison, we have plotted in Figure~\ref{fig:lognlogs} the $\log N-\log S$ relations for the real and random samples in the soft and hard bands, respectively. Both curves match very well along the entire flux range except at fluxes close to the flux limit of the survey, where the source counts of the simulated sample slightly underestimates the source counts of the real sample. However, the overall agreement is excellent and therefore the above method provides a good random sample for clustering analysis purposes.

The errors in different angular bins are not independent from one another. To estimate the errors we have followed the method described in Miyaji et al. (\cite{Miyaji07}) who computed the covariance matrix in the form

\begin{equation}
\label{eq:mcov}
\begin{array}{lcl}
M_{ij} & = & \sum_{k=1}^{N_{sim}}\left[w_R^k(\theta_i)-\langle w_R(\theta_i)\rangle\right]\left[w_R^k(\theta_j)-\langle w_R(\theta_j)\rangle\right]/N_{sim}\times \\
 & &  \times \left[1+w(\theta_i)\right]^{1/2}\left[1+w(\theta_j)\right]^{1/2}, \\
\end{array}
\end{equation}

\noindent where $w_R^k(\theta_i)$ is the angular correlation function for the $k$-th simulation in the $i$-th angular bin, $\langle w_R(\theta_i)\rangle$ is their mean value, $w(\theta_i)$ is the actual value for the angular correlation function computed from equation~\ref{eq:estim} and $N_{sim}$ is the total number of simulations. The square root of the diagonal elements $M_{ii}$ are hence the scaled errors for each bin in the angular correlation function.

\subsection{Fit to an analytical model}
\label{fits}

%-----------------------------------------
% theta-flux soft, hard
\begin{figure*}
\centering
\hbox{
\includegraphics[width=6.5cm,angle=270.0]{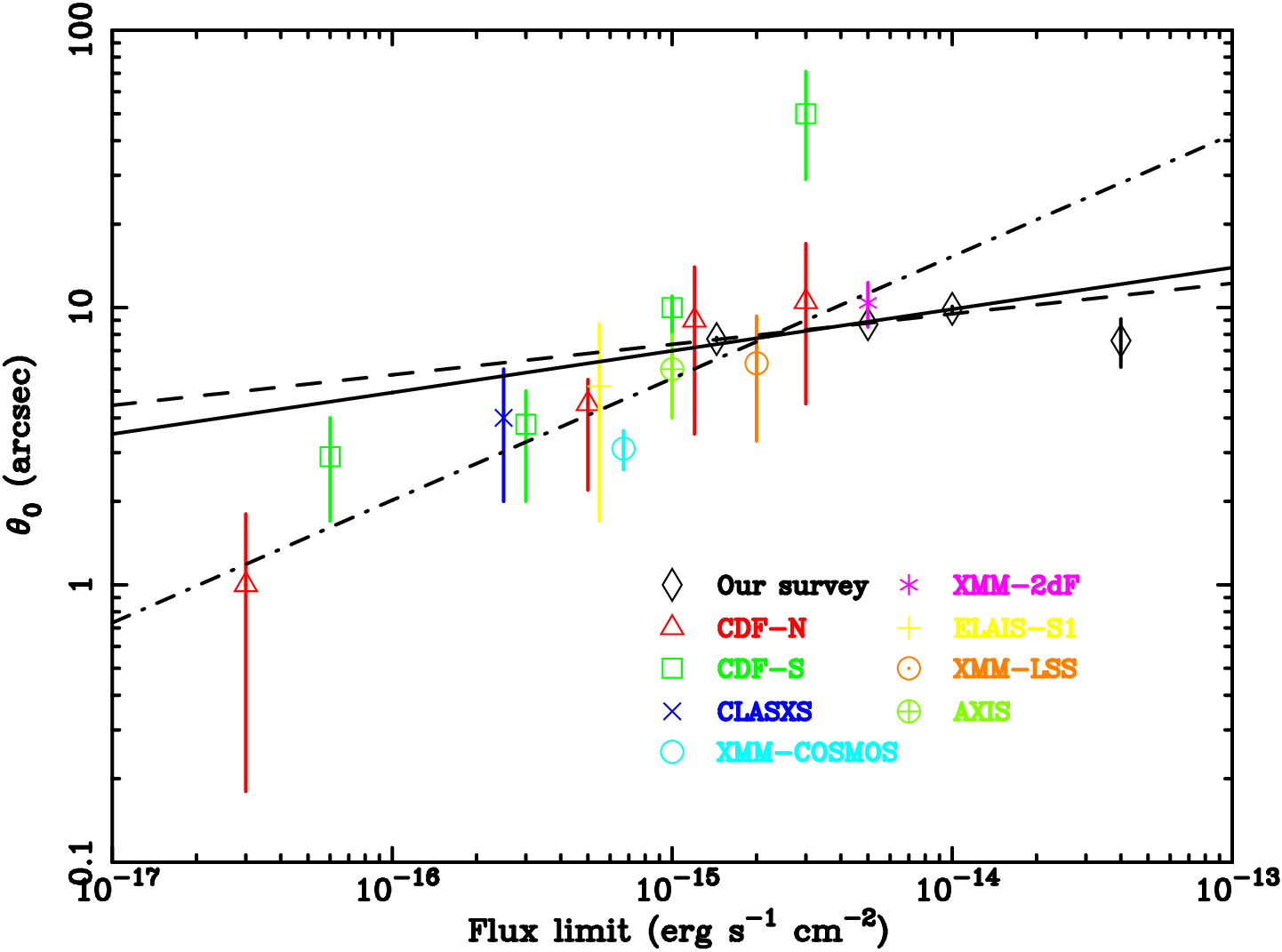}
\includegraphics[width=6.5cm,angle=270.0]{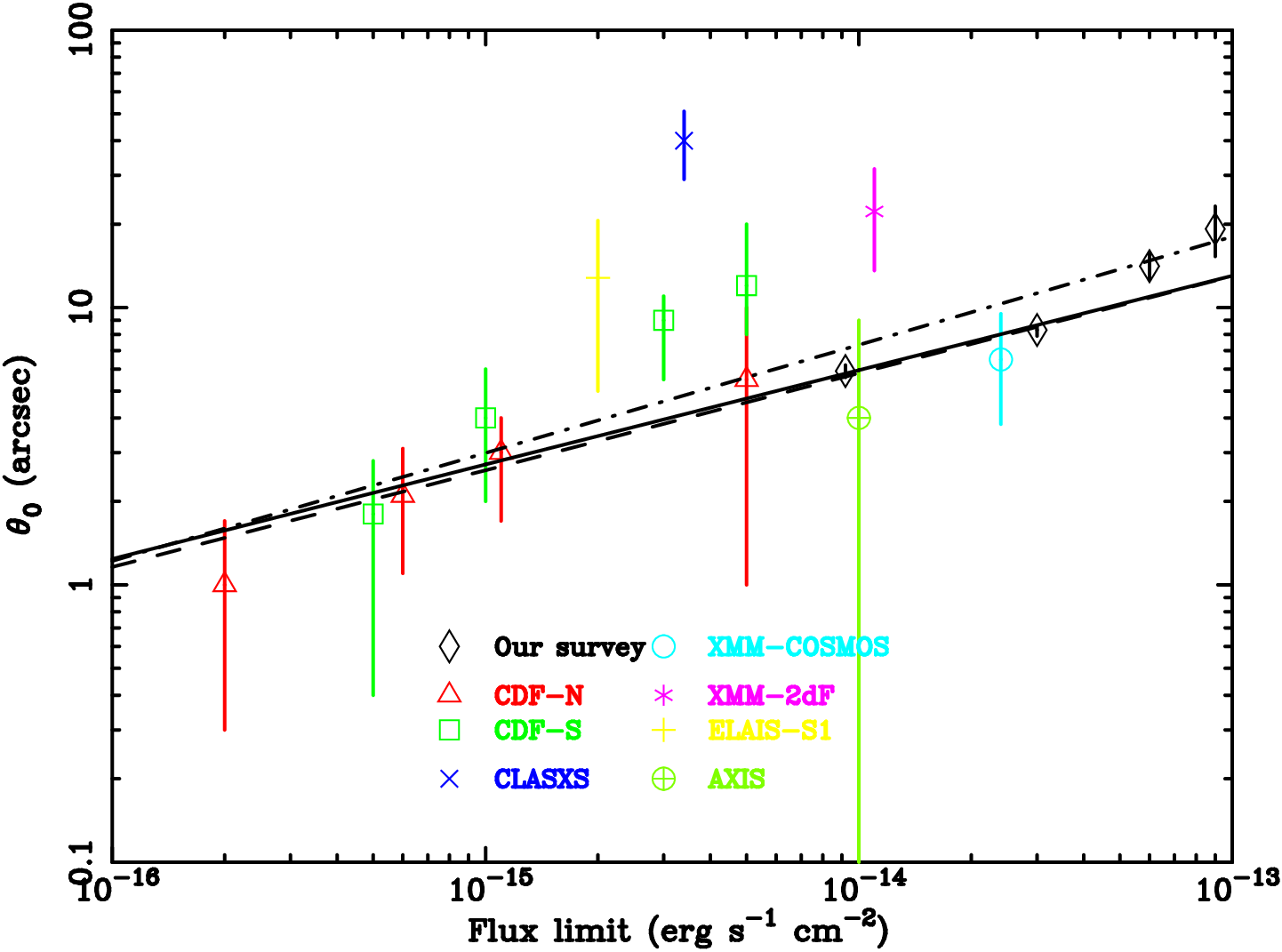}
}
\caption[]{Best-fit correlation length (for $\gamma-1 \equiv 0.8$) as a function of the flux limit of the sample (at zero area) in the soft ({\it Left panel}) and hard ({\it Right panel}) bands. Overplotted is the best fit to all points ({\it solid line}), our sample only ({\it dashed line}), and the rest of the surveys excluding our data points ({\it dot-dashed line}). The points belonging to our sample are not independent from each other (i.e. the sources are used cumulatively above the given flux limit).}
\label{fig:thetaflux}
\end{figure*}
%-------------------------------------------

The angular correlation function calculated in section~\ref{method} can be described by a power-law model in the form

\begin{equation}
\label{eq:model}
w_{model}(\theta)=\left(\frac{\theta}{\theta_0}\right)^{1-\gamma},
\end{equation}

\noindent where $1-\gamma$ is the slope and $\theta_0$ is the angular correlation length.

We have fitted the data using a $\chi^2$ technique. The fits were carried out over the range in which we had positive signal (50-1000 arcsec). Similarly as in Miyaji et al. (\cite{Miyaji07}), in order to take into account the correlations between errors computed in section~\ref{method} we have minimised the expression

\begin{equation}
\label{eq:chi2}
\chi^2 = \Delta^T M^{-1} \Delta,
\end{equation}

\noindent where $M^{-1}$ is the inverse of the covariance matrix (equation~\ref{eq:mcov}), and $\Delta$ is a vector such as

\begin{equation}
\label{eq:delta}
\Delta_i = w(\theta_i)-w_{model}(\theta_i)+IC.
\end{equation}

\noindent $IC$ is a constant that accounts for the integral constraint, which is a bias in the correlation function that occurs when a positive correlation is present at angular scales comparable to the individual field size. The mean surface density of objects from the survey is therefore too high thus producing a negative bias in the angular correlation function (Basilakos et al. \cite{Basilakos04}). The integral constraint can be formally calculated by

\begin{equation}
\label{eq:ic}
IC = \frac{\int \int d\Omega_1 d\Omega_2 w(\theta)}{\int \int d\Omega_1 d\Omega_2},
\end{equation}

\noindent where the integrals are carried out over the whole area of the survey. However, since the dependence of the sensitivity on the area of the survey is rather complicated we have estimated $IC$ empirically. For this, we have calculated the angular correlation function in the absence of correlation via the average of $N_{sim}$ realisations of $w(\theta)$ in which we have replaced the real data by random samples simulated independently using the method described in section~\ref{method} (Carrera et al. \cite{Carrera07}). The values of $IC$ thus obtained are small ($\sim 9.5 \times 10^{-3}$) but not negligible, and ignoring them can produce underestimations in the strength of the clustering signal.

\subsection{Results}
\label{results}

In this section we present the results obtained after fitting the overall samples to the power-law model described in section~\ref{fits} in each energy band. The best-fit parameters are summarised in Table~\ref{tab:fits} along with the number of sources involved. These values correspond to the fits in the 50-1000~arcsec range applying the integral constraint correction. The reported errors are 1$\sigma$. Fit results are plotted along with the observed binned angular correlation function in Figure~\ref{fig:wtheta}.

In the soft band (0.5-2~keV) we detect a high-significance ($\sim$10$\sigma$) clustering signal with a correlation length of $\theta_0=22.9\pm2.0$~arcsec and a slope of $\gamma-1=1.12\pm0.04$ after correcting for the integral constraint. If we ignore that correction in our fits, the correlation lengths are comparable within the error bars but the power-law becomes significantly steeper ($\gamma-1=1.29\pm0.04$). Similar values for the slope $\gamma-1$ in this band were found by Gandhi et al. (\cite{Gandhi06}) and Carrera et al. (\cite{Carrera07}), who obtained $\gamma-1=1.2$ albeit with larger (up to a factor 10) error bars. Our best-fit correlation length $\theta_0$ is consistent with that of Carrera et al. (\cite{Carrera07}) ($\theta_0=19_{-8}^{+7}$~arcsec) within the error bars, but it is significantly larger than that of Gandhi et al. (\cite{Gandhi06}) ($\theta_0=6.3\pm3.0$~arcsec). In general, our results without applying the integral constraint correction show slightly larger correlation lengths and steeper slopes, although generally consistent with the IC corrected results within the error bars.

%-------------------------
% Table of dependent-flux fits
\begin{table}
  \centering
  \caption[]{Summary of the fits in different flux-limited subsamples.}
  \label{tab:fluxfits}

  \begin{tabular}{lcccc}
    \hline
    \hline
    \noalign{\smallskip}
    Band (keV) &   Flux limit & $N_{sou}$ & $\theta_0$   & $\gamma-1$      \\
               &   (\cgs{}) &          &  (arcsec)     \\
    \noalign{\smallskip}
    \hline
    \noalign{\smallskip}
    Soft (0.5-2)   &   $5 \times 10^{-15}$ & 23264 & 27.3$_{-3.3}^{+3.2}$ & 1.16$_{-0.06}^{+0.05}$  \\
%       &   &  &  & 30.2$_{-3.3}^{+3.1}$ & 1.32$\pm$0.06  &  No  \\
       &    &  & 8.7$\pm$0.1 & 0.8 $(fixed)$   \\
%       &   &  &  & 6.3$\pm$0.1 & 0.8 $(fixed)$  &  No  \\
    \hline
    \noalign{\smallskip}
    Soft (0.5-2)   &   $1 \times 10^{-14}$ & 12046 & 37.3$_{-5.4}^{+5.0}$ & 1.28$_{-0.09}^{+0.08}$  \\
%       &   &  &  & 40.1$_{-5.2}^{+4.7}$ & 1.44$_{-0.10}^{+0.09}$  &  No  \\
       &    &  & 9.9$\pm$0.2 & 0.8 $(fixed)$   \\
%       &   &  &  & 7.3$\pm$0.2 & 0.8 $(fixed)$  &  No  \\
    \hline
    \noalign{\smallskip}
    Soft (0.5-2)    & $4 \times 10^{-14}$ & 1346 & 87.8$_{-13.4}^{+8.5}$ & 2.34$_{-0.34}^{+0.31}$   \\
%       &   &  &  & 85.7$_{-12.1}^{+7.7}$ & 2.61$_{-0.39}^{+0.38}$  &  No  \\
       &    &  & 7.6$\pm$1.5 & 0.8 $(fixed)$   \\
%       &   &  &  & 5.2$_{-1.3}^{+1.4}$ & 0.8 $(fixed)$  &  No  \\
    \hline
    \noalign{\smallskip}
    Hard (2-10)   &  $3 \times 10^{-14}$ & 4790 & 23.5$_{-12.4}^{+11.5}$ & 1.11$_{-0.24}^{+0.20}$   \\
%       &   &  &  & 26.1$_{-13.1}^{+11.4}$ & 1.26$_{-0.27}^{+0.23}$  &  No  \\
       &     &  & 8.3$_{-0.4}^{0.3}$ & 0.8 $(fixed)$   \\
%       &   &  &  & 6.0$\pm$0.3 & 0.8 $(fixed)$  &  No  \\
    \hline
    \noalign{\smallskip}
%    Hard (2-10)   &  50-1000 & $6 \times 10^{-14}$ & 1571 & 6.6$^a$ & 0.65$^a$  &  Yes  \\
%       &   &  &  & 9.0$^a$ & 0.75$^a$  &  No  \\
     Hard (2-10)  &  $6 \times 10^{-14}$  & 1571 & 14.1$_{-1.4}^{+1.5}$ & 0.8 $(fixed)$   \\
%       &   &  &  & 11.4$\pm$1.4 & 0.8 $(fixed)$  &  No  \\
    \hline
    \noalign{\smallskip}
    Hard (2-10)   &  $9 \times 10^{-14}$ & 609 & 19.2$_{-3.9}^{+4.0}$ & 0.8 $(fixed)$  \\
%       &   &  &  & 16.3$_{-3.7}^{+3.9}$ & 0.8 $(fixed)$  &  No  \\
    \hline
    \noalign{\smallskip}
%    \multicolumn{4}{l}{$^a$ Parameter unconstrained by the fit}
 \end{tabular}
\end{table}
%------------------------------------------

For comparison with other works that reported their results with a fixed $\gamma-1$ we have also made the fits fixing the slope to the canonical value of $\gamma-1=0.8$ (e.g. Peebles \cite{Peebles80}). The correlation length drops to $\theta_0=7.7\pm0.1$~arcsec then, which is in good agreement with Puccetti et al. (\cite{Puccetti06}) ($\theta_0=5.2\pm3.8$~arcsec) and Carrera et al. (\cite{Carrera07}) ($\theta_0=6\pm2$~arcsec) although our parameters are much more constrained thanks to the large statistics. Our result is smaller (by 2$\sigma$) than that computed by Basilakos et al. (\cite{Basilakos05}) using the \XMM{}/2dF survey ($\theta_0=10.4\pm1.9$~arcsec). On the other hand, Ueda et al. (\cite{Ueda08}) found a correlation length of $\theta_0=5.9_{-0.9}^{+1.0}$~arcsec in the SXDS survey, smaller than ours by $\sim$2$\sigma$, while Miyaji et al. (\cite{Miyaji07}) obtained a much weaker clustering strength of $\theta_0=3.1\pm0.5$~arcsec for sources in the COSMOS field. Similarly, Yang et at. (\cite{Yang03}) found a $\sim$2$\sigma$ clustering signal in \Chan{} observations of the Lockman Hole North-West region with $\theta_0=4\pm2$~arcsec. The samples used by Yang et al. (\cite{Yang03}) and Miyaji et al. (\cite{Miyaji07}) reach limiting fluxes one order of magnitude deeper than that of our sample. Therefore, their very low correlation lengths compared with ours could be explained in terms of a dependence of the clustering strength on the flux limit of the sample involved (see Section~\ref{fluxlimit}).

In the hard band (2-10~keV) the power-law becomes steeper ($\gamma-1=1.33$) and the clustering is stronger ($\theta_0=29.2_{-5.7}^{+5.1}$~arcsec) although marginally consistent with the results in the soft band within the 1$\sigma$ error bars. The clustering detection is still very significant ($\sim$5$\sigma$). The sources detected in the hard band are less biased against absorption, and if the unified model of AGN is correct (the obscuration of the nucleus is due to orientation effects only) one might not expect significant differences in the clustering properties of obscured and unobscured sources. However, accurate angular clustering measurements in this band have been difficult because of the limitations caused by the small-number statistics. For instance, Gandhi et al. (\cite{Gandhi06}), Carrera et al. (\cite{Carrera07}) and Ueda et al. (\cite{Ueda08}) were not able to obtain significant clustering signal in this band. Basilakos et al. (\cite{Basilakos04}) found results consistent with ours ($\gamma-1=1.2\pm0.3$ and $\theta_0=48.9_{-24.5}^{+15.8}$~arcsec) but with much larger error bars. Puccetti et al. (\cite{Puccetti06}) are also in agreement with us ($\theta_0=12.8\pm7.8$~arcsec) within their large uncertainties. For a canonical slope of 0.8, Yang et al. (\cite{Yang03}) found $\theta_0=40\pm11$~arcsec, much larger than ours ($\theta_0=5.9\pm0.3$~arcsec) with a significance of $\sim$4$\sigma$. Such a strong clustering is somewhat surprising and it is much higher than any other reported angular clustering analysis in this band using a fixed canonical slope (although it is roughly consistent with the result of Basilakos et al. \cite{Basilakos04} for a fixed $\gamma-1$=0.8). We would like to stress, however, that the characterisation of the angular correlation function using a fixed slope might not represent the true clustering of the X-ray sources (see the differences between both best-fit curves in Figure~\ref{fig:wtheta}). Indeed, the goodness of fit in both the soft and hard bands shows that when we allow the slope to float free, the fit to our measured $w(\theta)$ is better than when it is fixed to the canonical value of $\gamma=1.8$ (see Table~\ref{tab:fits}). However, we will make use of our fixed-slope fit results elsewhere in this paper in order to make comparisons with other works that only reported $\gamma=1.8$ results due to the low statistics.

We have also performed the angular clustering analysis in the ultrahard band (4.5-10~keV) obtaining inconclusive results, even after using only deeper fields (see Section~\ref{sample}). We found marginal clustering signal ($\sim$1$\sigma$) in this band, with the best-fit parameters consistent with those of the soft and hard bands (see Table~\ref{tab:fits}), and in good agreement with the results of Miyaji et al. (\cite{Miyaji07}).

\subsection{Dependence on the flux limit}
\label{fluxlimit}

Previous studies on the angular clustering of \Chan{} Deep Field sources using different subsamples suggest that the clustering strength might depend on the flux limit of the sample (Giacconi et al. \cite{Giacconi01}, Plionis et al. \cite{Plionis08}). Similar results with other samples with different depths seem to point out also in this direction. We have investigated this behavior in our soft and hard samples (where we have enough statistics) by splitting both samples into subsamples with different flux limits and then computing their angular correlation funtion. This means that we are using all sources cumulatively above the given flux limit. The results of these fits are reported in Table~\ref{tab:fluxfits}. Errors are 1$\sigma$.

We can see that if we leave both parameters $\theta_0$ and $\gamma-1$ free, the clustering strength significantly increases and the power law becomes steeper as we move towards brighter flux limits in the soft band. Something similar is observed in the hard band, although in this case we were unable to simultaneously fit both parameters in the brightest subsamples due to the low source density.

%-----------------------------------------
% theta-flux binned
\begin{figure}
\centering
\hbox{
\includegraphics[width=6.5cm,angle=270.0]{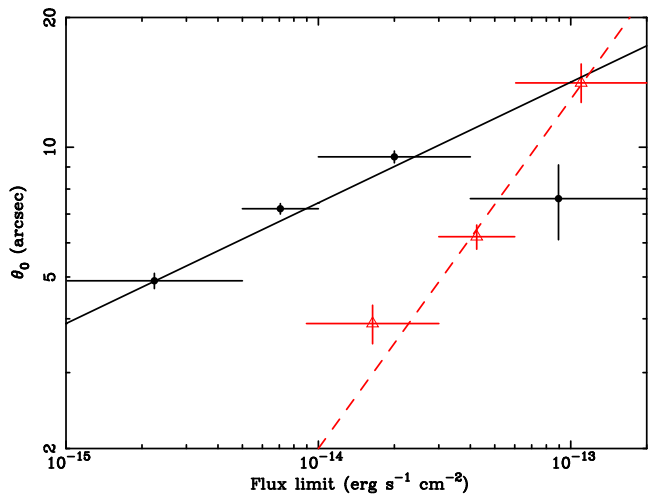}
}
\caption[]{Best-fit correlation length (for $\gamma-1 \equiv 0.8$) as a function of the flux limit of the sample in the soft ({\it dots}) and hard ({\it triangles}) bands in different flux bins. The individual points are independent from each other. Overplotted is the soft ({\it solid line}) and hard ({\it dashed line}) best fits.}
\label{fig:thetafluxbinned}
\end{figure}
%-------------------------------------------

%-----------------------------------------
% wtheta soft, hard, HR selected
\begin{figure*}
\centering
\hbox{
\includegraphics[width=6.5cm,angle=270.0]{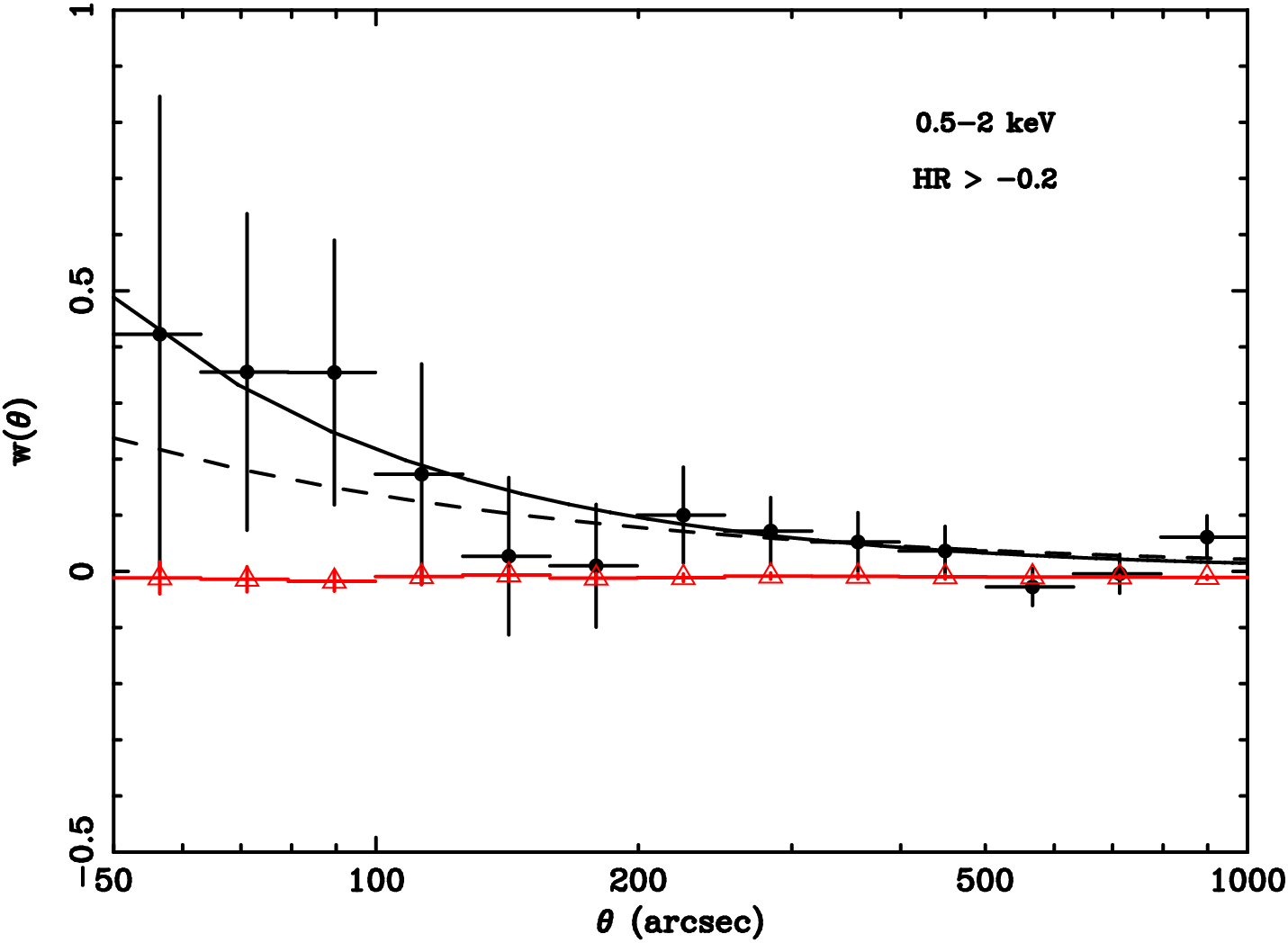}
\includegraphics[width=6.5cm,angle=270.0]{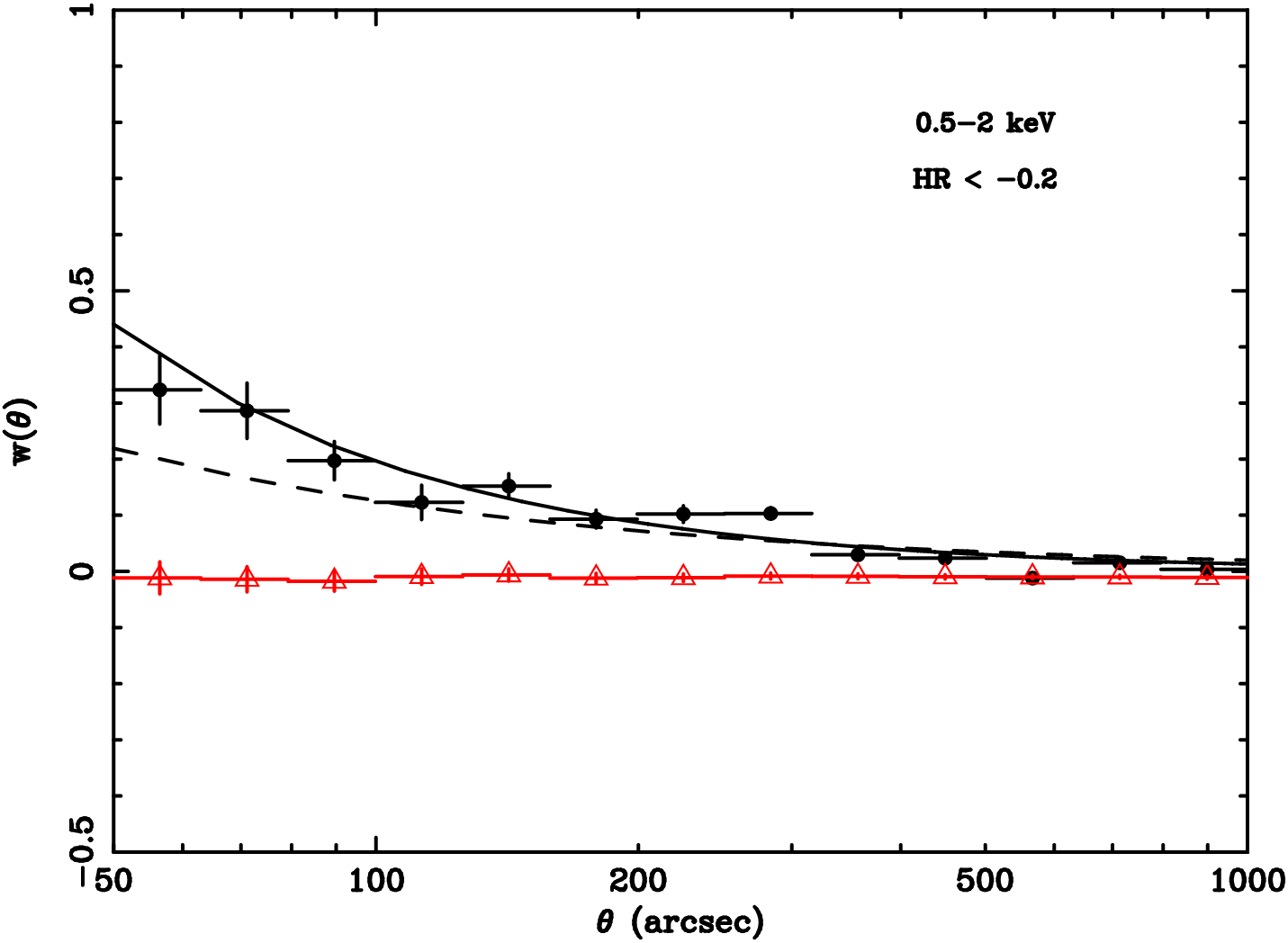}
}
\hbox{
\includegraphics[width=6.5cm,angle=270.0]{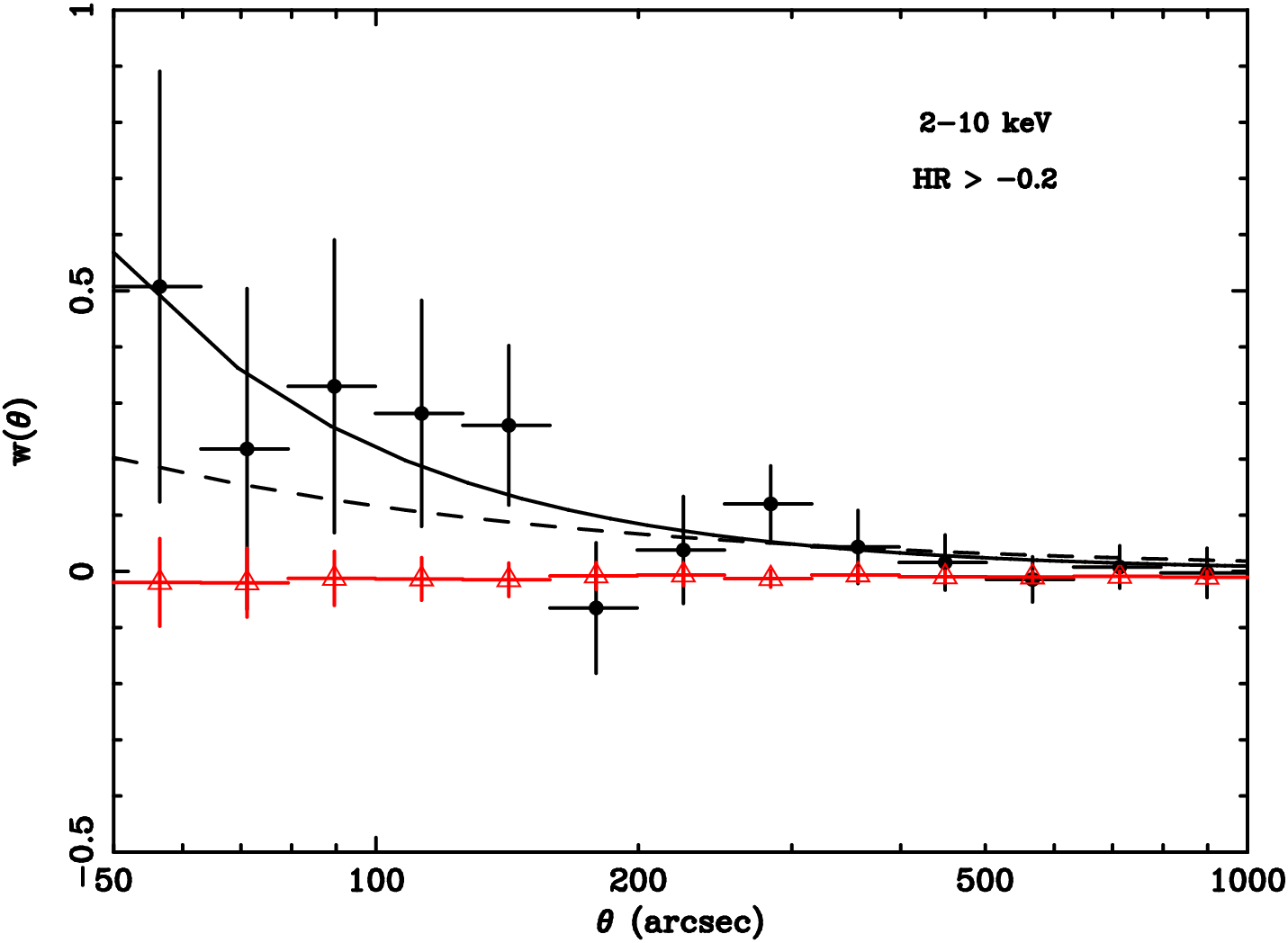}
\includegraphics[width=6.5cm,angle=270.0]{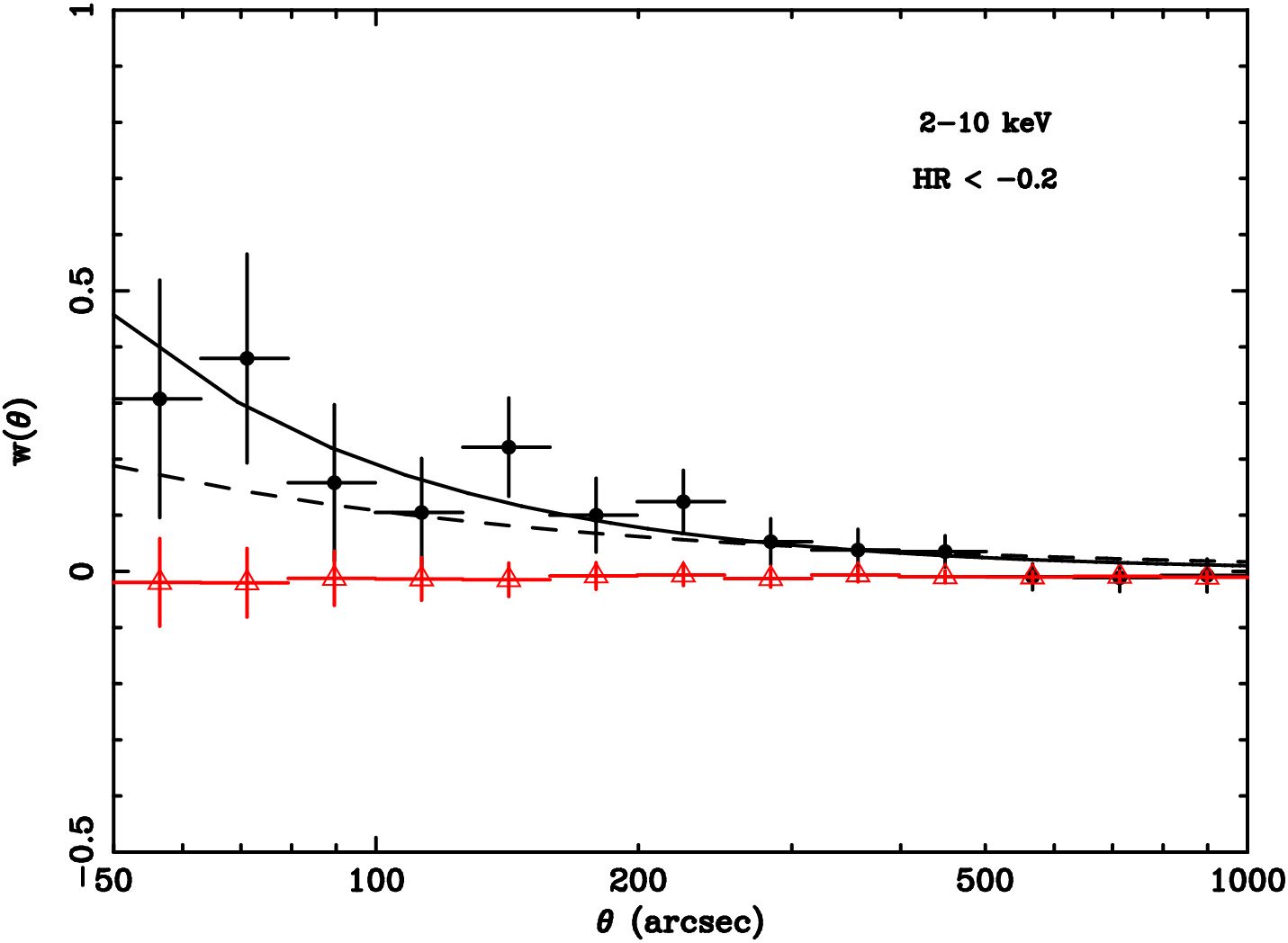}
}
\caption[]{Angular correlation function for the soft $HR\geq -0.2$ ({\it upper left panel}), soft $HR < -0.2$ ({\it upper right panel}), hard $HR \geq -0.2$ ({\it lower left panel}), and hard $HR < -0.2$ ({\it lower right panel}) sources. Solid dots are the observed data. Solid triangles represent the average random estimation of $w(\theta)$ used to calculate the integral constraint (see text). Overplotted is the best-fit $\chi^2$ with and without fixed slope (dashed and solid lines, respectively).}
\label{fig:wthetaHR}
\end{figure*}
%-------------------------------------------

In order to compare these results with other works we froze $\gamma-1 \equiv 0.8$. The resulting best-fit correlation length $\theta_0$ still shows a strong dependence with the flux limit of the subsample. In Figure~\ref{fig:thetaflux} we have plotted these points along with the results from other surveys such as CDF-N and CDF-S (Plionis et al. \cite{Plionis08}), CLASXS (Yang et al. \cite{Yang03}), COSMOS (Miyaji et al. \cite{Miyaji07}), XMM-2dF (Basilakos et al. \cite{Basilakos04}, \cite{Basilakos05}), ELAIS-S1 (Puccetti et al. \cite{Puccetti06}), XMM-LSS (Gandhi et al. \cite{Gandhi06}) and AXIS (Carrera et al. \cite{Carrera07}). In Figure~\ref{fig:thetaflux} there seems to be a linear trend $\theta_0$-flux in spite of the large uncertainties in most of the surveys.

We tried to formalize the observed trend by fitting the points to a power-law in the form $\theta_0 = T(S/S_{15})^\alpha$, where $S_{15}= 1 \times 10^{-15}$~\cgs{} and the normalisation $T$ is the expected correlation length at $S_{15}$. We find that $\alpha = 0.15\pm0.01$ and $T = 6.98\pm0.10$~arcsec in the soft band, and $\alpha = 0.34\pm0.04$ and $T = 2.72_{-0.29}^{+0.31}$~arcsec in the hard band, respectively. 

Although our points match well with the results from other surveys within the error bars of the latter, they seem to follow a slightly different trend. We therefore fitted our points alone obtaining $\alpha = 0.11\pm0.01$ and $T = 7.37_{-0.12}^{+0.11}$~arcsec, and $\alpha = 0.35_{-0.04}^{+0.06}$ and $T = 2.59_{-0.39}^{+0.41}$~arcsec in the soft and hard bands, respectively. Fitting the rest of the surveys data points and excluding ours provide the following best-fit parameters: $\alpha = 0.44_{-0.07}^{+0.08}$ and $T = 5.55\pm0.39$~arcsec in the soft band, and $\alpha = 0.39_{-0.08}^{+0.07}$ and $T = 2.99_{-0.45}^{+0.46}$~arcsec in the hard band.

There are no significant differences between the fits in the hard band, all of them consistent with each other. In the soft band, however, strong differences arise, our data points suggesting a much milder dependence on the flux limit than that predicted by other surveys. Although our data points do not look visually misplaced with respect to the general trend (with the exception of the last point at the highest flux limit, which could be severely affected by the low statistics), the fits in which they are involved are dominated by them due to their much smaller error bars. We have also checked that the above results do not significantly change by removing the point at the highest flux limit, the fits still dominated by the tiny error bars of the rest of the points. On the other hand, the \Chan{} Deep Field data points alone suggest a stronger dependence in both energy bands with respect to the one we would expect from the other surveys, which might be caused by cosmic variance or the so-called amplification bias (caused by the merging of a pair of sources into a single source due to the PSF of the detector at small angular separations, Vikhlinin \& Forman \cite{Vikhlinin95}) as discussed in Gilli et al. (\cite{Gilli05}) and Plionis et al. (\cite{Plionis08}).

We have also checked whether the flux limit dependence is still significant using independent flux bins. Our data points plotted in Figure~\ref{fig:thetaflux} are not independent from each other, since the sources above a given flux limit are also present in the brighter subsamples. Therefore, we have studied the flux dependence using independent data points. As shown in Figure~\ref{fig:thetafluxbinned} the trend is also very clear in this case, being much steeper in the hard band than in the soft band ($\alpha=0.67\pm0.09$ against $\alpha=0.26\pm0.02$) thus suggesting that the clustering strength of the sources detected in the 2-10~keV band is significantly more dependent on the flux limit of the subsample.

There are several possible explanations for the $\theta_0$-flux limit trend. One of them could be that, since X-ray surveys at different flux limits are, in principle, sampling different populations of sources, they have different clustering properties. Deep pencil-beam surveys will probe fainter and typically further sources than wide-area bright surveys, thus suggesting a possible dependence of clustering with redshift. Other simple explanation could be that pairs of sources in deep surveys (hence with fainter flux limits) tend to be separated at smaller angular distances due a projection effect. If this is the actual cause of the trend, the dependence should disappear after taking into account the real separation between sources (i.e. redshifts).

\subsection{Angular clustering of a hardness ratio selected sample}
\label{HRangular}

If the unified model of AGN is correct and the obscuration is due to an orientation effect, the clustering of obscured and unobscured sources should not be significantly different. The expected fraction of obscured AGN in our sample is $\sim$50\% in the 2-10~keV band, and $\sim$20\% in the 0.5-2~keV band (Mateos et al. \cite{Mateos08}), and they can be selected by computing their hardness ratios ($HR$)

\begin{equation}
\label{eq:hr}
HR=\frac{H-S}{H+S},
\end{equation}

\noindent where $S$ and $H$ are the count rate of the source in the soft and hard bands, respectively. A $HR$ value often used in the literature to discriminate between obscured and unobscured sources is $HR=-0.2$, which approximately corresponds to a source with an obscuring column density $N_H=10^{22}$~cm$^{-2}$ and an intrinsic power-law photon index of 1.7 at redshift $z=0.7$ (see e.g. Gandhi et al. \cite{Gandhi04}).

We have applied this criterion to our sources thus creating two subsamples with $HR\geq-0.2$ and $HR<-0.2$ in the soft and hard bands, and studied their angular clustering properties. The best-fit parameters are reported in Table~\ref{tab:HRfits} along with their 1$\sigma$ uncertainties. The results are plotted in Figure~\ref{fig:wthetaHR}.

The results show that the clustering properties of $HR \geq -0.2$ sources, which are likely to be absorbed, are not significantly different from that of the $HR < -0.2$ sources (well within the error bars) in both the soft and hard bands. This is in contradiction with the results of Gandhi et al. (\cite{Gandhi06}) who found that sources in the \XMM{}-LSS survey with $HR>-0.2$ in the 2-10~keV band showed clustering evidence whereas sources with $HR<-0.2$ did not. They were hugely affected by low-count statistics, however, and their angular clustering detections were marginal.

On the other hand, our results are in agreement with those of Gilli et al. (\cite{Gilli05}) and (\cite{Gilli09}), who computed the spatial correlation function for obscured and unobscured sources in the CDF and COSMOS fields, respectively. They were unable to find significant evidence of different clustering behaviours between both subsets of sources. Moreover, since obscured AGN are typically detected at low redshifts ($z<1$) in medium-depth surveys, this could also mean that sources above and below redshift 1, when AGN reached a maximum in comoving density in the Universe according to recent luminosity function models, do not cluster differently.

%-------------------------
% Table of HR fits
\begin{table}
  \centering
  \caption[]{Summary of the fits of the hardness-ratio selected subsamples.}
  \label{tab:HRfits}

  \begin{tabular}{lccc}
    \hline
    \hline
    \noalign{\smallskip}
    Band (keV) &   $N_{sou}$ & $\theta_0$   & $\gamma-1$     \\
               &             &  (arcsec)     \\
    \noalign{\smallskip}
    \hline
    \noalign{\smallskip}
    Soft (0.5-2) $HR\geq-0.2$  &   4038 & 27.1$_{-17.7}^{+14.5}$ & 1.17$_{-0.35}^{+0.26}$   \\
%       &   &   & 31.6$_{-18.0}^{+13.4}$ & 1.37$_{-0.37}^{+0.28}$  &  No  \\
       &    & 8.3$\pm$0.5 & 0.8 $(fixed)$    \\
%       &   &   & 5.9$\pm$0.4 & 0.8 $(fixed)$  &  No  \\
    \hline
    \noalign{\smallskip}
    Soft (0.5-2) $HR<-0.2$  &    27791 & 24.8$_{-2.7}^{+2.6}$ & 1.17$\pm$0.05   \\
%       &   &   & 27.4$_{-2.7}^{+2.6}$ & 1.34$_{-0.06}^{+0.05}$ &  No  \\
       &    & 7.5$\pm$0.1 & 0.8 $(fixed)$    \\
%       &   &   & 5.2$\pm$0.1 & 0.8 $(fixed)$  &  No  \\
    \hline
    \noalign{\smallskip}
    Hard (2-10) $HR\geq-0.2$ &   3425  & 33.1$_{-17.8}^{+13.3}$ & 1.37$_{-0.36}^{+0.28}$    \\
%      &  &   & 35.8$_{-17.7}^{+12.4}$ & 1.57$_{-0.40}^{+0.31}$  & No   \\
      &    & 6.8$\pm$0.5 & 0.8 $(fixed)$    \\
%      &  &   & 4.7$\pm$0.5 & 0.8 $(fixed)$  & No   \\
    \hline
    \noalign{\smallskip}
    Hard (2-10) $HR<-0.2$ &  6006  & 27.0$_{-9.9}^{+8.5}$ & 1.27$_{-0.20}^{+0.17}$    \\
%      &  &   & 29.2$_{-10.1}^{+8.3}$ & 1.46$_{-0.23}^{+0.19}$  & No   \\
      &    & 6.2$\pm$0.3 & 0.8 $(fixed)$    \\
%      &  &   & 4.1$\pm$0.3 & 0.8 $(fixed)$  & No   \\
    \hline
    \noalign{\smallskip}
  \end{tabular}
\end{table}
%------------------------------------------

%_________________________________________________

\section{Spatial clustering}
\label{spatial}

\subsection{Inversion of Limber's equation}
\label{limber}

The two-dimensional angular correlation function is a projection in the sky of the real three-dimensional spatial correlation function $\xi(r)$ along the line of sight, where $r$ is the physical separation between sources (typically in units of $h^{-1}$~Mpc).

We can model the spatial correlation function as (De Zotti et al. \cite{DeZotti90})

\begin{equation}
\xi(r,z)=\left(\frac{r}{r_0}\right)^{-\gamma}(1+z)^{-(3+\epsilon)},
\label{xi}
\end{equation}

\noindent where $\epsilon$ parameterizes the type of clustering evolution. For instance, if $\epsilon=\gamma - 3$, the clustering is constant in comoving coordinates, which means that the amplitude of the correlation function remains fixed with redshift in comoving coordinates as the pair of sources expands together with the Universe. On the other hand, if $\epsilon=-3$ the clustering is constant in physical coordinates (De Zotti et al. \cite{DeZotti90}).

%--------------------------------------------------
\begin{figure}
\centering
\includegraphics[width=6.5cm,angle=270.0]{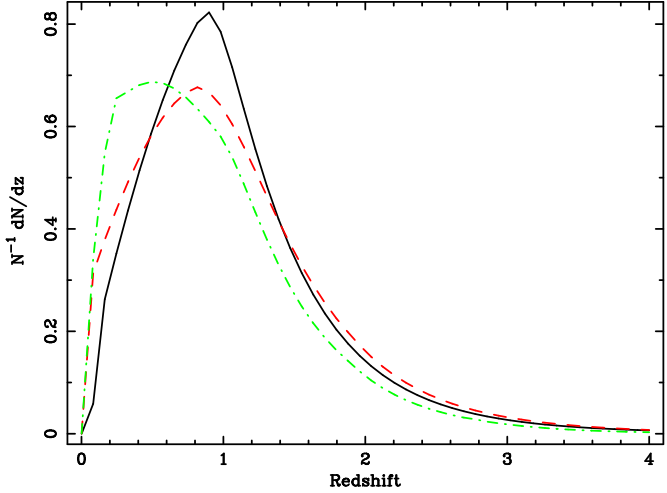}
\caption[]{The redshift selection function for the soft ({\it solid line}), hard ({\it dashed line}) and ultrahard ({\it dot-dashed line}) bands.}
\label{fig:selfunc}
\end{figure}
%---------------------------------------------------

The angular amplitude $\theta_0$ can be related to the spatial amplitude $r_0$ by inverting Limber's integral equation (Peebles \cite{Peebles93}). In the case of a spatially flat Universe, Limber's equation can be expressed as (Basilakos et al. \cite{Basilakos05})

\begin{equation}
w(\theta)=2\frac{\int_0^{\infty}\int_0^{\infty}D_C^4\phi^2(D_c)\xi(r,z)dD_cdu}{\left(\int_0^\infty D_c^2\phi(D_c)dD_c\right)^2},
\label{limbereq}
\end{equation}

\noindent where $\phi(D_c)$ is the selection function (the probability that a source at a comoving distance $D_c$ is detected in the survey). The comoving distance $D_c$ is related to the redshift through (Hogg \cite{Hogg99})

\begin{equation}
D_c(z)=\frac{c}{H_0}\int_0^z\frac{dz'}{E(z')},
\label{Dc}
\end{equation}

\noindent with

\begin{equation}
E(z)=\left[\Omega_m(1+z)^3+\Omega_\Lambda\right]^{1/2}
\label{Ez}
\end{equation}

\noindent in a spatially flat Universe ($\Omega_k=0$), and $H_0$ being the Hubble constant (Peebles \cite{Peebles93}, Hogg \cite{Hogg99}).

The number of objects in a survey that subtend a solid angle $\Omega_S$ in the sky within a redshift shell $dz$ is

\begin{equation}
\frac{dN}{dz}=\Omega_SD_c^2\phi(D_c)\left(\frac{c}{H_0}\right)E^{-1}(z).
\label{dNdz}
\end{equation}

Combining these equations, Limber's equation becomes

\begin{equation}
w(\theta)=2\frac{H_0}{c}\int_0^\infty\left(\frac{1}{N}\frac{dN}{dz}\right)^2E(z)dz\int_0^\infty\xi(r,z)du.
\label{limbereq2}
\end{equation}

The physical separation between two sources that project an angle $\theta$ in the sky can be written as

\begin{equation}
r \simeq \frac{1}{1+z}(u^2+x^2\theta^2)^{1/2}
\label{raprox}
\end{equation}

\noindent under the small angle approximation assumption. If we now combine equations~\ref{xi} and \ref{limbereq2} we obtain

\begin{equation}
\theta_0^{\gamma-1}=H_\gamma\left(\frac{r_0^\gamma H_0}{c}\right)\int_0^\infty\left(\frac{1}{N}\frac{dN}{dz}\right)^2\frac{E(z)(1+z)^{-3-\epsilon+\gamma}}{D_c^{\gamma-1}(z)}dz,
\label{limbereq3}
\end{equation}

\noindent where

\begin{equation}
H_\gamma=\frac{\Gamma(\frac{1}{2})\Gamma(\frac{\gamma-1}{2})}{\Gamma(\frac{\gamma}{2})},
\label{Hgamma}
\end{equation}

\noindent with $\Gamma$ being the gamma function.

We can now invert equation~\ref{limbereq3} to obtain $r_0$ given an angular amplitude $\theta_0$, but we also need to determine the source redshift distribution $dN/dz$, which can be estimated from a given luminosity function model. We can do this via the equation~\ref{dNdz} writing the selection function (degradation of sampling as a function of the distance in flux-limited surveys) as

\begin{equation}
\phi(D_c)=\int_{L_{min}(z)}^\infty\Phi(L_x,z)dL_x,
\label{phi}
\end{equation}

\noindent which depends on the evolution of the source luminosity function $\Phi(L_x,z)$ but is independent of the cosmological model.

We have determined the redshift selection function (see Figure~\ref{fig:selfunc}) for our sample using the best-fit luminosity-dependent density evolution (LDDE) model of the X-ray luminosity function of Ebrero et al. (\cite{Ebrero08}) for all three bands (soft, hard and ultrahard). Our results on the inversion of Limber's equation for different values of the clustering evolution parameter $\epsilon$ are listed in Table~\ref{tb:limber}, along with the median of the redshift distribution in each band. The angular clustering parameters $\theta_0$ and $\gamma$ used are those of Table~\ref{tab:fits} corrected from the integral constraint. The 1$\sigma$ errors of $r_0$ have been computed assuming fixed $\gamma$ and $\epsilon$.

Although it is an useful tool to estimate the spatial clustering, the inversion of Limber's equation has some limitations since a number of assumptions has to be done. The spatial correlation $r_0$ obtained this way is therefore affected by uncertainties coming from the determination of $\theta_0$ and $\gamma$, the type of clustering evolution $\epsilon$, and the redshift selection function. In this context, we can assume that the values of the angular correlation function $\theta_0$ and $\gamma$ obtained in this work have been accurately determined thanks to the large statistics involved (at least in the soft and hard bands). The parameter $\epsilon$ is hardly known in any system and therefore all our results will be reported for both clustering models.

A critical issue for the inversion of Limber's equation is the redshift selection function $N^{-1}dN/dz$. The luminosity function model from which it is derived was computed using AGN from a variety of public surveys ranging from deep-pencil beam to shallow wide surveys, and the intrinsic absorption was taken into account at hard X-rays (Ebrero et al. \cite{Ebrero08}). Hence, we can assume that the redshift selection function is robust enough within the redshift range in which it was computed ($z \sim 0-3$).

%----------------------------------
%
\begin{table}
  \centering
   \caption[]{Spatial correlation lengths $r_0$ for different clustering models ($\epsilon$) obtained from Limber's equation. Errors are 1$\sigma$.}
\label{tb:limber}
   \begin{tabular}{lcccc}
       \hline
       \noalign{\smallskip}
Band (label)  &  $\gamma$ & $\epsilon$ & $r_0$ ($h^{-1}$~Mpc) & $\bar{z}$ \\
       \noalign{\smallskip}
       \hline
      \noalign{\smallskip}
Soft (S1) &   2.12  &  -0.88  & $12.25\pm0.12$ & 0.96 \\
Soft (S2) &   2.12  &  -3     & $6.54\pm0.06$  & 0.96 \\
Soft (S3) &   1.80  &  -1.20  & $13.74\pm0.14$ & 0.96 \\
Soft (S4) &   1.80  &  -3     & $7.20\pm0.07$  & 0.96 \\
       \noalign{\smallskip}
       \hline
Hard (H1) &   2.33  &  -0.67  & $9.9\pm2.4$  & 0.94 \\
Hard (H2) &   2.33  &  -3     & $5.7\pm1.4$  & 0.94 \\
Hard (H3) &   1.80  &  -1.20  & $12.7\pm0.5$ & 0.94 \\
Hard (H4) &   1.80  &  -3     & $6.8\pm0.3$  & 0.94 \\
       \noalign{\smallskip}
       \hline
Ultrahard (U1) &   2.47  &  -0.53  & $7.0\pm5.5$  & 0.77 \\
Ultrahard (U2) &   2.47  &  -3     & $5.1\pm4.1$  & 0.77 \\
Ultrahard (U3) &   1.80  &  -1.20  & $11.6\pm1.7$ & 0.77 \\
Ultrahard (U4) &   1.80  &  -3     & $7.4\pm1.1$  & 0.77 \\
      \noalign{\smallskip}
       \hline
   \end{tabular}
\end{table}
%
%-----------------------------------

Our results in the soft band assuming clustering constant in physical coordinates ($\epsilon=-3$) are in good agreement with those obtained from optically selected AGN surveys $r_0 \simeq 5.4-8.6 h^{-1}$~Mpc (Akylas et al. \cite{Akylas00}, Croom et al. \cite{Croom02}, Grazian et al. \cite{Grazian04}) and X-ray selected surveys such as Mullis et al. (\cite{Mullis04}) ($r_0=7.4_{-1.9}^{+1.8} h^{-1}$~Mpc) or Basilakos et al. (\cite{Basilakos05}) ($r_0=7.5\pm0.6 h^{-1}$~Mpc). However, when we assume evolution constant in comoving coordinates ($\epsilon=\gamma-3$) the result of Basilakos et al. (\cite{Basilakos05}) is slightly larger (at 2$\sigma$ confidence level) while the predictions of Miyaji et al. (\cite{Miyaji07}) are systematically smaller ($r_0 \simeq 9.8\pm0.7 h^{-1}$~Mpc).

In the hard band, our deprojected spatial amplitude is significantly smaller than the value from Basilakos et al. (\cite{Basilakos04}) who obtained $r_0\simeq12-19 h^{-1}$~Mpc using hard sources from the \XMM{}-2dF survey for a fixed canonical slope $\gamma=1.8$. However, they find that their correlation lengths are much larger (over a factor 2) than the ones provided in the literature for AGN (Croom et al. \cite{Croom01}) or 2dF (Hawkins et al. \cite{Hawkins03}) and SDSS (Budavari et al. \cite{Budavari03}) galaxy distributions. In fact, the $r_0$ values obtained by Basilakos et al. (\cite{Basilakos04}) can be compared instead to that of extremely red objects (EROs) and luminous radio sources (Roche et al. \cite{Roche03}, Overzier et al. \cite{Overzier03}, R\"ottgering et al. \cite{Rottgering03}) which are in the range $r_0\simeq12-15 h^{-1}$~Mpc. On the other hand, the results of Basilakos et al. (\cite{Basilakos04}) for a best-fit slope $\gamma=2.2$ are consistent with ours within the error bars, which may imply that the fixed slope representation is not a good characterisation for the large-scale clustering of AGN.

Gilli et al. (\cite{Gilli05}) performed a broadband spatial clustering analysis of the X-ray sources detected in the \Chan{} Deep Field North and South obtaining correlation lengths in the range $r_0=5-10 h^{-1}$~Mpc, roughly in agreement with our predictions within the error bars, albeit with a much flatter slope $\gamma \sim 1.4$. More recently, Gilli et al. (\cite{Gilli09}) studied the spatial clustering of AGN in the COSMOS field finding $r_0=8.65_{-0.48}^{+0.41} h^{-1}$~Mpc and $\gamma=1.88$.

In the context of a comoving clustering scenario $\epsilon=\gamma-3$, these results in the soft and hard bands seem to confirm the difference in the spatial clustering between X-ray selected and optically selected AGN reported in other works. Our large correlation length $r_0 > 10$$h^{-1}$~Mpc is in general consistent with that of X-ray selected AGN (Basilakos et al. \cite{Basilakos04}, Basilakos et al. \cite{Basilakos05}, Puccetti et al. \cite{Puccetti06}), while reported correlation lengths from optically selected AGN are significantly shorter ($r_0\simeq5$$h^{-1}$~Mpc, Croom et al. \cite{Croom02}). The situation turns when we assume clustering in physical coordinates $\epsilon=-3$, being our computed $r_0$ consistent with the values of Croom et al. (\cite{Croom02}).

Our results in the ultrahard band are in good agreement (within the error bars) with the deprojected spatial clustering calculated by Miyaji et al. (\cite{Miyaji07}) using COSMOS sources detected in the 4.5-10~keV range. It must be taken into account, however, that the $\theta_0$ values we used to invert Limber's equation in this band correspond only to a marginal detection of clustering.

%-----------------------------------------
% r0 versus Lx, z
\begin{figure*}
\centering
\hbox{
\includegraphics[width=6.5cm,angle=270.0]{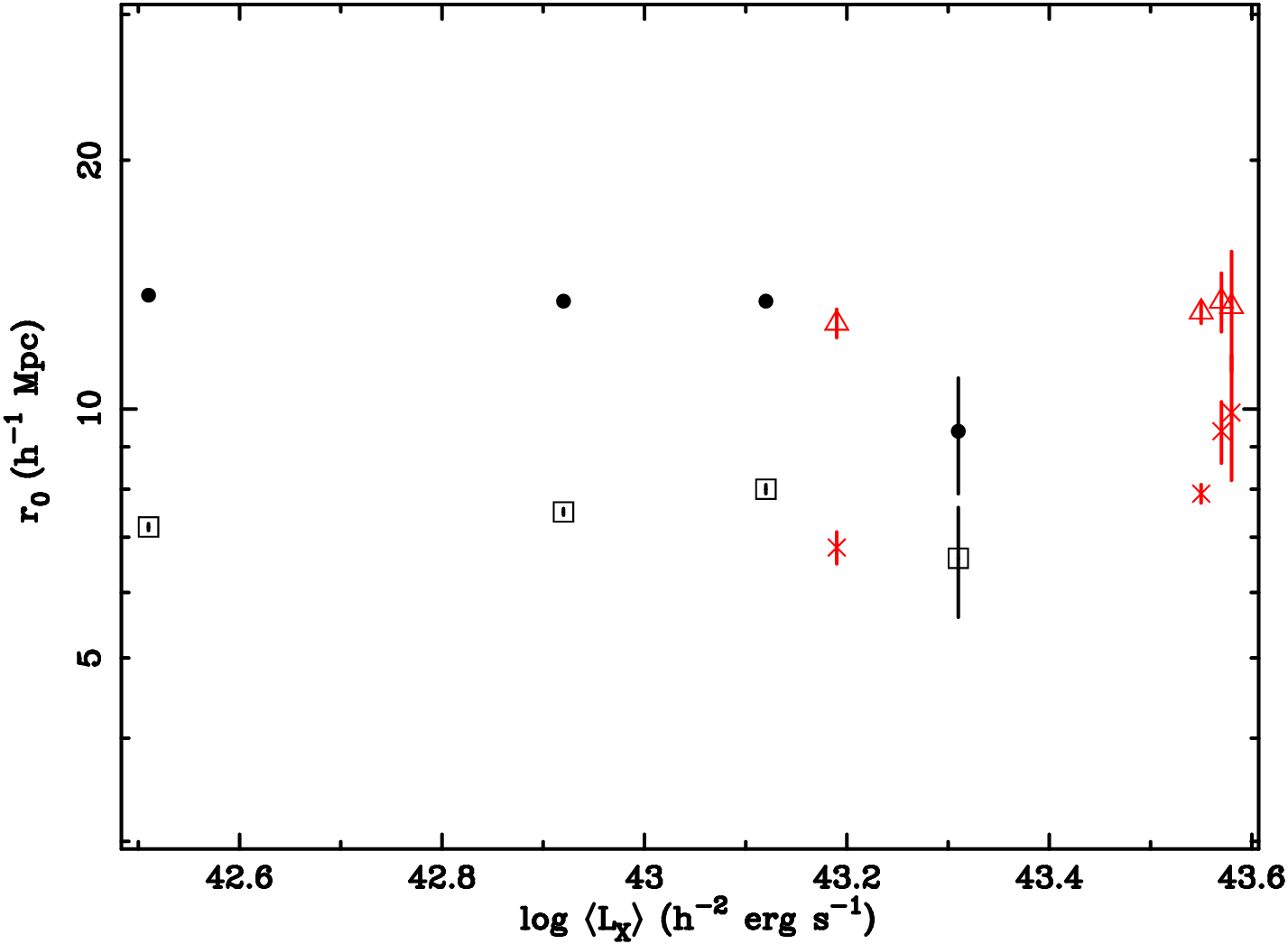}
\includegraphics[width=6.5cm,angle=270.0]{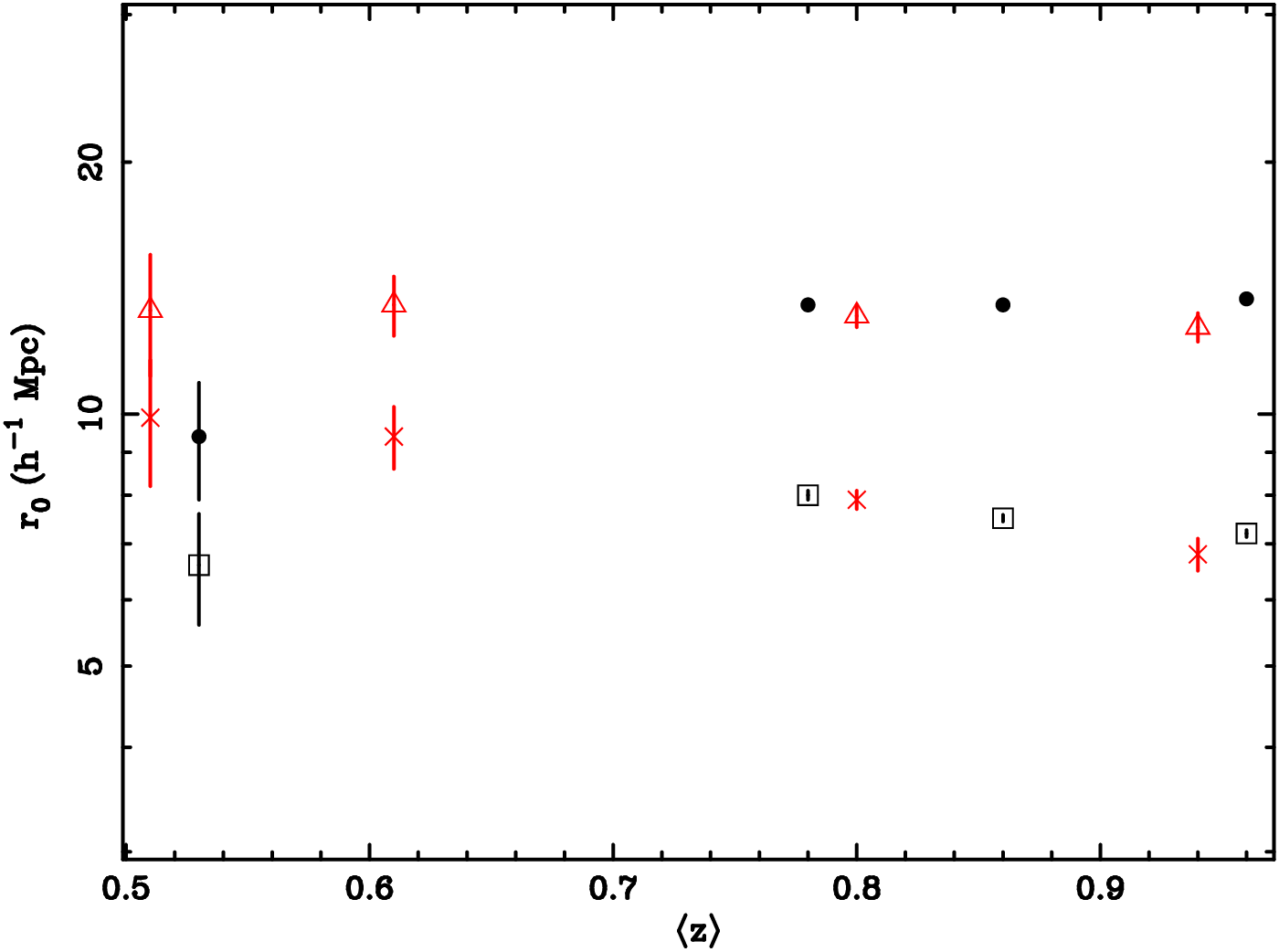}
}
\caption[]{Deprojected spatial correlation length $r_0$ as a function of the median luminosity ({\it left panel}) and redshift ({\it right panel}). The results are shown for different clustering models $\epsilon=-1.2$ (soft band: dots; hard band: triangles) and $\epsilon=-3$ (soft: squares; hard: crosses).}
\label{fig:rlz}
\end{figure*}
%-------------------------------------------

\subsection{Dependence on the X-ray luminosity and redshift}
\label{luminosity}

The results obtained in Section~\ref{fluxlimit} showed that the angular clustering strength seemed to depend on the flux limit of the sample under consideration. Since there are several combinations of redshifts and luminosities that yield to a given flux, we have investigated whether the deprojected spatial length $r_0$ depends on the X-ray luminosity and redshifts. Moreover, this would provide additional information on the evolution of the X-ray sources.

Since flux and luminosity are related to each other through the luminosity distance $S=L/4\pi d^2_L(z)$ and $\theta_0\propto r_0^{\gamma/\gamma-1}$ (via Limber's equation), if we assume that the $\theta_0-S$ relation described in Section~\ref{fluxlimit} $\theta_0\propto S^\alpha$ is generally true, we would expect a spatial correlation length dependent on luminosity and redshift in the form $r_0\propto (L/d^2_L(z))^{\alpha(\gamma-1)/\gamma}$.

We have therefore inverted Limber's equation for different flux-limited subsamples (see Section~\ref{fluxlimit}). The median redshifts and luminosities have been derived from the LDDE best-fit luminosity function of Ebrero et al. (\cite{Ebrero08}). In Figure~\ref{fig:rlz} we plot the deprojected correlation distances $r_0$ as a function of the median luminosity and redshift of the different subsamples for different clustering models.

We find no significant dependence of the clustering strength neither on the luminosity nor on the redshift median ranges spanned by our sample. In a comoving clustering scenario, there seem to be no dependence on the median luminosity in neither the soft and hard bands (dots and triangles in the left panel of Figure~\ref{fig:rlz}, respectively). However, when we consider $\epsilon=-3$ a slightly positive trend is observed in both bands (squares and crosses, respectively, same panel). In order to check out the significance of this dependency, we fitted both a constant value and a constant plus a linear term to the points in Figure~\ref{fig:rlz} and compared their goodnesses of fit. The F-test results show no significant improvement of the fit for neither model and therefore we conclude that no significant dependence of $r_0$ on luminosity is found in our data. On the other hand, Plionis et al. (\cite{Plionis08}) found indications for a luminosity dependent clustering in the \Chan{} Deep Fields, although their large error bars make this conclusion rather uncertain. Using our much more constrained best-fit parameters we are unable to confirm such dependency.

Similarly, we do not see evolution on the clustering properties of our sources, although a mild dependence could be present as $r_0$ slightly decreases as the median redshift increases. The spatial clustering analysis of the \Chan{} Deep Field (Gilli et al. \cite{Gilli05}) and COSMOS sources (Gilli et al. \cite{Gilli09}) led to similar conclusions.

\subsection{Bias parameter and connection to dark matter haloes}
\label{bias}

The spatial clustering values obtained in section~\ref{limber} can be used to estimate the mass of the dark matter haloes (DMH) in which these sources are embedded. A commonly used quantity for such an analysis is the bias parameter, that is usually defined as

\begin{equation}
\label{eq:biasdef}
b^2(z)=\frac{\xi_{AGN}(8,z)}{\xi_{DMH}(8,z)},
\end{equation}

\noindent where $\xi_{AGN}(8,z)$ and $\xi_{DMH}(8,z)$ are the spatial correlation functions of AGN and DMH evaluated at 8 $h^{-1}$~Mpc, respectively. The former value has been calculated in this work whereas the latter can be estimaded using (Peebles \cite{Peebles80})

\begin{equation}
\label{eq:sigma8}
\sigma_8^2(z)=\xi(8,z)J_2,
\end{equation}

\noindent where $J_2=72/\left[(3-\gamma)(4-\gamma)(6-\gamma)2^\gamma\right]$ and $\sigma_8^2(z)$ is the dark matter density variance in a sphere with a comoving radius of 8 $h^{-1}$~Mpc which evolves as

\begin{equation}
\label{eq:sigma8z}
\sigma_8(z)=\sigma_8D(z).
\end{equation}

\noindent $D(z)$ is the linear growth factor of perturbations, which is defined as $D_{EdS}(z)=(1+z)^{-1}$ in an Einstein-De Sitter cosmology. However, in the context of a $\Lambda$-CDM cosmology the growth of perturbations is weaker and it is attenuated by a suppression factor $g(z)$ so that $D(z)=g(z)(1+z)^{-1}$. We have used here the analytical approximation for $g(z)$ described in Carroll et al. (\cite{Carroll92}). On the other hand, we have fixed the $rms$ dark matter fluctuation at present time $\sigma_8$ to 0.84 (Spergel et al. \cite{Spergel03}).

The CDM structure formation scenario predicts that the bias parameter is determined by the dark matter halo mass (Mo \& White \cite{Mo96}). We have used the large-scale bias relation as a function of halo mass reported in Sheth et al. (\cite{Sheth01})

\begin{equation}
\label{eq:bias}
\begin{array}{lcl}
b(M,z) & = & 1+\frac{1}{\sqrt{a}\delta_c(z)} \times\\
       &   & \left[a\nu^2\sqrt{a}+0.5\sqrt{a}(a\nu^2)^{1-c}-\frac{(a\nu^2)^c}{(a\nu^2)^c+0.5(1-c)(1-c/2)}\right], \\
\end{array}
\end{equation}

\noindent where $\nu=\delta_c(z)/\sigma(M,z)$, $a=0.707$, and $c=0.6$. $\delta_c$ is the critical overdensity for collapse of a homogeneous spherical perturbation, and it takes the value of $\simeq1.69$ in an Einstein-De Sitter cosmology. For a general cosmology, $\delta_c$ posseses a weak dependence on redshift as reported in Navarro et al. (\cite{Navarro97}). $\sigma(M,z)$ is the $rms$ density fluctuation in the linear density field that evolves

\begin{equation}
\label{eq:sigmaMz}
\sigma(M,z)=\sigma(M)D(z)
\end{equation}

\noindent where $\sigma(M)$ is given by the convolution of a power spectrum $P(k)$ with a top-hat window function $w(k)$,

\begin{equation}
\label{eq:sigmaM}
\sigma^2(M)=\frac{1}{2\pi^2}\int_0^\infty k^2P(k)w^2(k)dk.
\end{equation}

\noindent For a power-law power spectrum $P(k)\propto k^n$, the $rms$ fluctuation on mass is

\begin{equation}
\label{eq:sigmaMpl}
\sigma(M)=\sigma_8\left(\frac{M}{M_8}\right)^{-(n+3)/6},
\end{equation}

\noindent where $M_8$ is the characteristic mean mass within 8 $h^{-1}$~Mpc (see e.g. Martini \& Weinberg \cite{Martini01}).

%----------------------------------
%
\begin{table}
  \centering
   \caption[]{Bias and dark matter halo mass. Errors are 1$\sigma$.}
\label{tb:bias}
   \begin{tabular}{lcccc}
       \hline
       \noalign{\smallskip}
Band (label)$^a$  &  $\sigma_8$ & b & $\log M_{DMH}$$^b$  \\
       \noalign{\smallskip}
       \hline
      \noalign{\smallskip}
Soft (S1) &   2.52$\pm$0.10  &  4.82$\pm$0.18  & 12.76$\pm$0.31  \\
Soft (S2) &   1.30$\pm$0.03  &  2.48$\pm$0.07  & 12.50$\pm$0.31   \\
Soft (S3) &   2.22$\pm$0.01  &  4.24$\pm$0.02  & 12.71$\pm$0.30  \\
Soft (S4) &   1.24$\pm$0.01  &  2.37$\pm$0.01  & 12.49$\pm$0.30   \\
       \noalign{\smallskip}
       \hline
Hard (H1) &   2.39$\pm$0.56  &  4.53$\pm$1.06  & 12.74$\pm$0.34   \\
Hard (H2) &   1.26$\pm$0.27  &  2.38$\pm$0.51  & 12.49$\pm$0.34   \\
Hard (H3) &   2.07$\pm$0.06  &  3.92$\pm$0.11  & 12.69$\pm$0.31  \\
Hard (H4) &   1.18$\pm$0.03  &  2.23$\pm$0.06  & 12.47$\pm$0.31   \\
       \noalign{\smallskip}
       \hline
Ultrahard (U1) &   1.81$\pm$1.70  &  3.16$\pm$2.97  & 12.66$\pm$0.44   \\
Ultrahard (U2) &   1.22$\pm$1.08  &  2.14$\pm$1.88  & 12.50$\pm$0.44   \\
Ultrahard (U3) &   1.91$\pm$0.15  &  3.34$\pm$0.26  & 12.68$\pm$0.32  \\
Ultrahard (U4) &   1.27$\pm$0.09  &  2.23$\pm$0.16  & 12.52$\pm$0.32   \\
      \noalign{\smallskip}
       \hline
       \multicolumn{4}{l}{$^a$ Labels are those of the fits in Table~\ref{tb:limber}} \\
       \multicolumn{4}{l}{$^b$ In units of $h^{-1}$~M$_\odot$} \\
   \end{tabular}
\end{table}
%
%-----------------------------------

The results for the bias parameters and estimated DMH masses for the different clustering models in the soft, hard and ultrahard bands are reported in Table~\ref{tb:bias}. We find an average $\langle \log M_{DMH}\rangle=12.50\pm0.34$ $h^{-1}$~M$_\odot$ and $\langle \log M_{DMH}\rangle=12.71\pm0.34$ $h^{-1}$~M$_\odot$ for the clustering models $\epsilon=-3$ and $\epsilon=\gamma-3$, respectively. This is in excellent agreement with the results from the AERQS survey (Grazian et al. \cite{Grazian04}) or the 2dF survey (Porciani et al. \cite{Porciani04}, Croom et al. \cite{Croom05}) and, more recently, the COSMOS survey (Gilli et al. \cite{Gilli09}).

Our derived bias parameters $b$ are in excellent agreement with those reported in Basilakos et al. (\cite{Basilakos08}), who found $b(z=1.2)=4.88\pm1.20$ and $b(z=0.85)=4.65\pm1.50$ in the soft and hard bands, respectively. In general, X-ray selected AGN show larger bias parameters than optically selected AGN (e.g. Croom et al. \cite{Croom05}, Myers et al. \cite{Myers07}). This could mean that the underlying matter distribution is traced differently in X-rays and in the optical domain, with X-ray selected AGN residing in more massive DMH than the optically selected AGN. A number of works (Porciani et al. \cite{Porciani04}, Croom et al. \cite{Croom05}, Basilakos et al. \cite{Basilakos08}) show that optical AGN are likely to be hosted by DMH with masses $\lesssim 10^{-13}$ $h^{-1}$~M$_\odot$, while X-ray AGN are usually embedded in DMH with masses $\gtrsim 10^{-13}$ $h^{-1}$~M$_\odot$. Our results, however, are slightly lower but consistent nevertheless with the estimations from the spatial clustering of COSMOS sources ($M_{DMH}\sim 12.4-12.8$ $h^{-1}$~M$_\odot$, Gilli et al. \cite{Gilli09}). We must stress that our results are derived from a mostly unidentified X-ray sample which has been selected so that the vast majority of the sources are likely to be AGN, although we can expect some pollution coming from Galactic stars and passive galaxies. We estimate the population of non-AGN sources in our sample to be of the order of $\sim$10\% (see e.g. Barcons et al. \cite{Barcons07}).

\begin{figure*}
\centering
\hbox{
\includegraphics[width=6.5cm,angle=270.0]{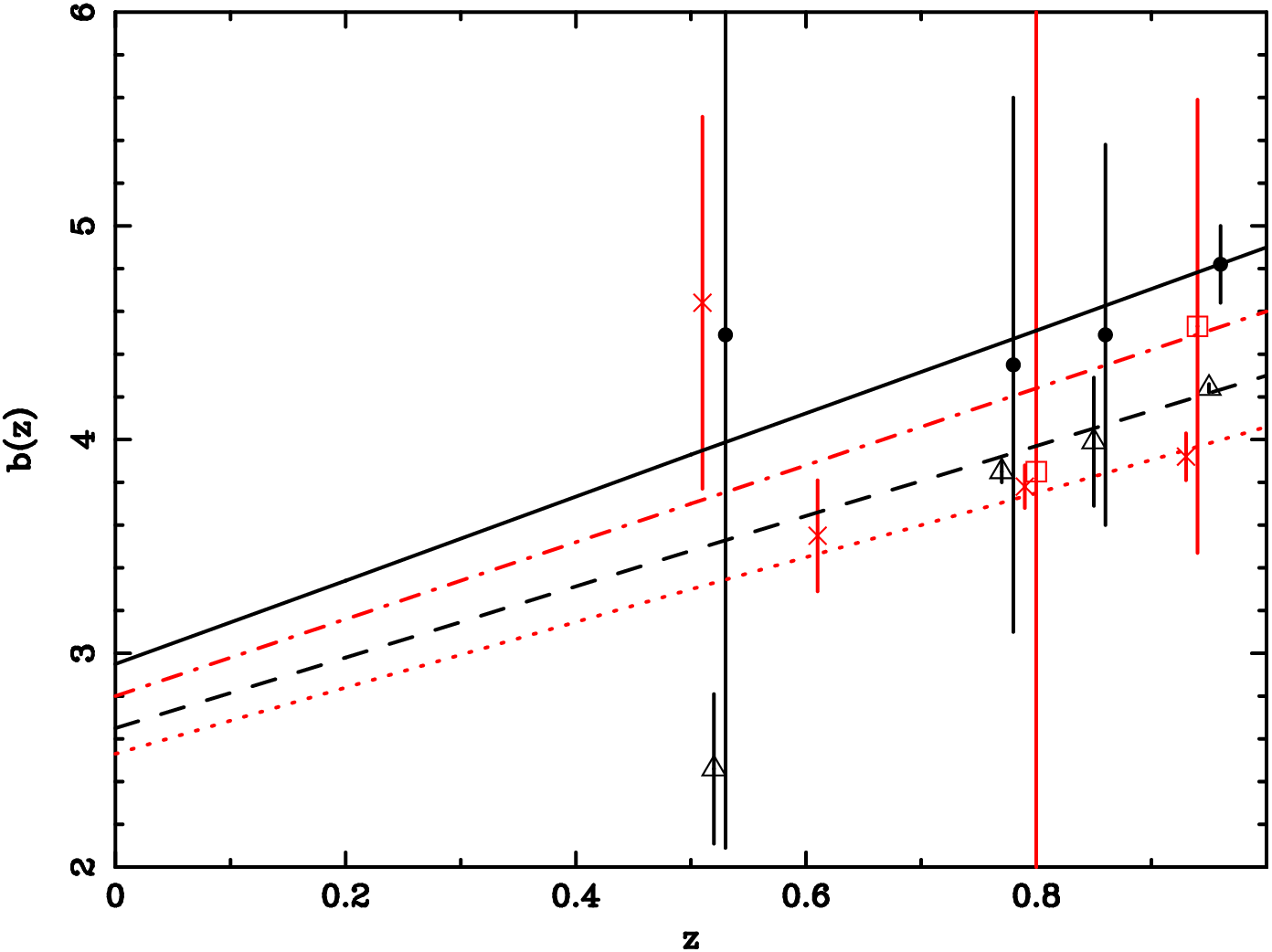}
\includegraphics[width=6.5cm,angle=270.0]{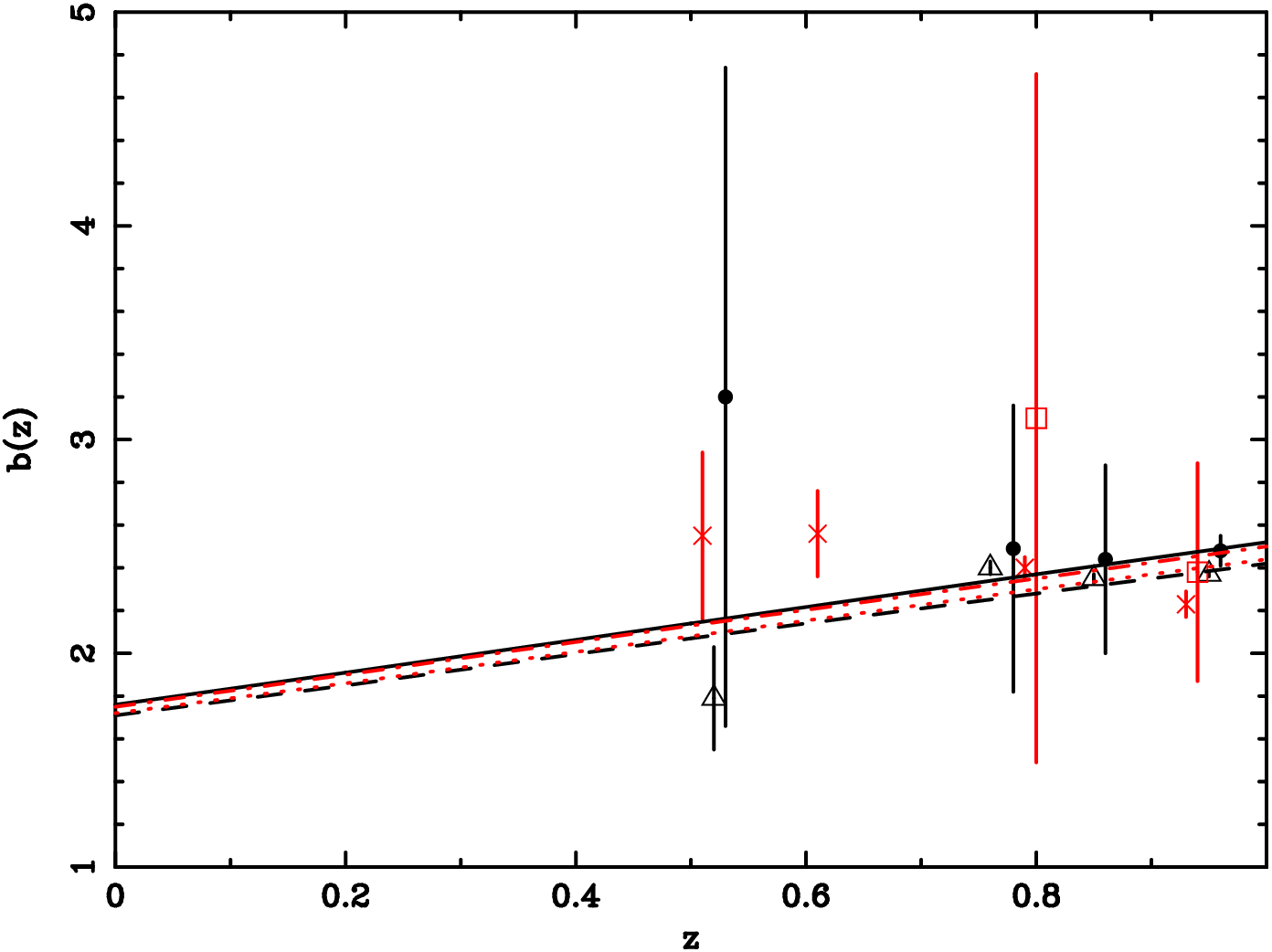}
}
\caption[]{Bias parameter as a function of redshift for different bands and clustering models. {\it Left panel:} Clustering model $\epsilon=\gamma-3$ fits in the soft (S1 and S3, dots and triangles, respectively), and hard (H1 and H3, squares and crosses, respectively) bands. {\it Right panel:} Clustering model $\epsilon=-3$ fits in the soft (S2 and S4, dots and triangles, respectively), and hard (H2 and H4, squares and crosses, respectively) bands. Overplotted are the best fits to the {\it conserving} bias evolution model in each band.}
\label{fig:bz}
\end{figure*}

It is possible to trace the bias evolution with redshift using the relations described above. At redshifts up to $\sim$1, where the median of the redshift distribution of our sample is expected to lie, a simple model to describe the bias evolution is the so-called {\it conserving model} (Nusser \& Davis \cite{Nusser94}, Fry \cite{Fry96}),

\begin{equation}
\label{eq:bz}
b(z)=1+(b_0-1)/D(z),
\end{equation}

\noindent where $b_0$ is the population bias at $z=0$. This model assumes that the objects, after being formed at a given high-redshift epoch, evolve with time within the gravitational potential. We have computed the bias parameter for several subsamples with different median redshifts in the soft band, finding that the present-time bias $b_0$ strongly depends on the clustering model. For instance, for the $\epsilon=\gamma-3$ model (fits S1, S3, H1 and H3), $b_0$ lies in the range 2.5-3.0, whereas for the $\epsilon=-3$ model (fits S2, S4, H2, H4) we obtained $b_0 \sim 1.75$ (see Table~\ref{tb:b0} and Figure~\ref{fig:bz}). These results were expected since the $\epsilon=-3$ model removes the redshift dependence in equation~\ref{xi} and hence produces lower correlation lengths with respect to the $\epsilon=\gamma-3$ model. Similar results for $b_0$ were obtained by Basilakos et al. (\cite{Basilakos05}), who also studied the bias evolution for both clustering models. Gilli et al. (\cite{Gilli09}) found $b_0$ values in the range 1.5-2, similar to that of Croom et al. (\cite{Croom05}), consistent with our predictions for a $\epsilon=-3$ clustering scenario.

%-----------------------------------------
% b(z)
%-------------------------------------------

\subsection{The lifetime of AGN}
\label{lifetime}

We can estimate the lifetime of AGN using the mean DMH masses calculated above and making some simple assumptions. We have followed the method proposed by Martini \& Weinberg (\cite{Martini01}), assuming that we are sampling the most massive DMH at a given redshift $z$ and that each DMH hosts an active AGN at any given time. There is hence a relation between the comoving density of AGN $\Phi(z)$ and their lifetime $t_{AGN}(z)$ in the form

\begin{equation}
\label{ndens}
\Phi(z)=\int_{M_{min}}^\infty \frac{t_{AGN}(z)}{t_{DMH}(M,z)}n(M,z)dM,
\end{equation}

\noindent where $M_{min}$ is the minimum halo mass hosting an AGN, $n(M,z)$ is the comoving density of DMH of mass $M$ at redshift $z$, and $t_{DMH}(M,z)$ is the lifetime of DMH.

%----------------------------------
%
\begin{table}
  \centering
   \caption[]{Predicted bias at $z=0$.}
\label{tb:b0}
   \begin{tabular}{lcc}
       \hline
       \noalign{\smallskip}
Band (label)  &  $\epsilon$ & $b_0$  \\
       \noalign{\smallskip}
       \hline
      \noalign{\smallskip}
Soft (S1) &   $\gamma-3$  & $2.95\pm0.09$  \\
Soft (S2) &    -3     & $1.76\pm0.04$   \\
Soft (S3) &   $\gamma-3$  & $2.65\pm0.01$  \\
Soft (S4) &    -3     & $1.71\pm0.01$   \\
       \noalign{\smallskip}
       \hline
Hard (H1) &  $\gamma-3$  & $2.80\pm0.02$  \\
Hard (H2) &    -3     & $1.75\pm0.25$  \\
Hard (H3) &  $\gamma-3$  & $2.53\pm0.04$  \\
Hard (H4) &    -3     & $1.72\pm0.02$  \\
       \noalign{\smallskip}
       \hline
   \end{tabular}
\end{table}
%
%-----------------------------------

According to Martini \& Weinberg (\cite{Martini01}), the characteristic halo timelife is defined as the time interval during which a DMH of mass $M$ at redshift $z$ is incorporated to a larger halo of mass $2M$. To a first approximation, we can assume that the halo lifetime is comparable to the Hubble time at that redshift, $t_{DMH}(M,z) \sim t_U(z)$. Equation~\ref{ndens} then yields

\begin{equation}
\label{eq:lifetime}
t_{AGN}(z)=t_U(z)\frac{\Phi(z)}{\Phi_{DMH}(z)},
\end{equation}

\noindent where

\begin{equation}
\label{DMHdens}
\Phi_{DMH}(z)=\int_{M_{min}}^\infty n(M,z)dM
\end{equation}

\noindent is the comoving density of DMH with mass above $M_{min}$.

We can estimate $\Phi_{DMH}(z)$ following the Press-Schechter approximation as described in Martini \& Weinberg (\cite{Martini01}), obtaining a DMH comoving space density at $z=1$ (approximately the median of the redshift distribution of our sample) for haloes with mass larger than $\log M_{min} = 12.6$~$h^{-1}$~M$_\odot$ (the average halo mass estimated in section~\ref{bias}) of $\Phi_{DMH} \simeq 2 \times 10^{-3}$~$h^3$~Mpc$^{-3}$.

For the comoving density of AGN, we have used the predicted value from the X-ray luminosity function of Ebrero et al. (\cite{Ebrero08}) that we used to deproject Limber's equation, at the median luminosity of our sample and $z=1$. This yielded to an AGN duty cycle in the range $t_{AGN}/t_U = 0.054-0.078$. In our cosmological framework, the Hubble time at $z=1$ is $\sim$5.8~Gyr. The estimated lifetime of AGN is hence in the range $t_{AGN}=3.1-4.5 \times 10^8$~yr.

However, the assumptions made in the Martini \& Weinberg (\cite{Martini01}) approximation (i.e. AGN activity is a random event in the lifetime of a halo) might not be valid at redshifts below $z<2$, where the AGN space density begins to decline and hence fuelling mechanisms may trigger AGN activity rather than black hole growth, thus dominating clustering properties.

The estimated lifetime derived above corresponds to the total activity period for a single AGN, which can be split into several episodes of activity. The significantly shorter lifetime compared to the time period spanned between redshift $\sim$1 and 0 indicates that we are probably observing several generations of AGN, and that an important fraction of galaxies might experience AGN activity one or more times throughout their lifes.

%_________________________________________________

\section{Conclusions}
\label{conclusions}

We have studied the angular correlation function of a large sample of serendipitous X-ray sources from 1063 \XMM{} observations at high Galactic latitudes in several energy bands: 0.5-2 (soft), 2-10 (hard) and 4.5-10 (ultrahard) keV. Our sample comprises 31288, 9188 sources in the soft and hard bands, respectively, covering $\sim$125.5~deg$^{2}$ in the sky, and 1259 sources in the ultrahard band over $\sim$51.5~deg$^{2}$, thus being the largest sample ever used in clustering investigations.

We found significant positive angular clustering signal in the soft ($\sim$10$\sigma$) and hard ($\sim$5$\sigma$) bands, while our results in the ultrahard band are only marginal ($<$1$\sigma$). The result in the hard band clears up the debate on whether X-ray sources detected in this band cluster or not, since a number of past works had reported different inconclusive results ranging from a few $\sigma$ detections to no detection at all.

We made power-law fits to the angular correlation function taking into account correlations between errors in the range 50-1000~arcsec, determining the best-parameters with unprecedent accuracy. We obtained typical correlation lengths of $\theta_0=$22.9$\pm$2.0 (soft), $\theta_0=$29.2$_{-5.7}^{+5.1}$ (hard), and $\theta_0=$40.9$_{-29.3}^{+19.6}$ (ultrahard), and slopes $\gamma-1=$ of 1.12$\pm$0.04, 1.33$_{-0.11}^{+0.10}$, and 1.47$_{-0.57}^{+0.43}$, in the soft, hard and ultrahard bands, respectively. An angular clustering characterisation with a fixed canonical slope of $\gamma-1=1.8$, typical value found for nearby galaxies, does not reproduce well the observed data.

Previous angular clustering studies reported that the clustering strength might depend on the flux limit of the sample (Giacconi et al. \cite{Giacconi01}, Plionis et al. \cite{Plionis08}). Indeed, after splitting our sample into several subsamples at different flux limits we found a dependency of $\theta_0$ on the flux limit. One possible explanation for this behaviour is that different flux limits effectively sample different source populations, thus reflecting an underlying dependence of the clustering properties on redshift.

We have also studied the angular clustering of hardness-ratio selected subsamples in the soft and hards, finding that the clustering properties of sources with $HR \geq -0.2$ are not significantly different to that of $HR < -0.2$ sources. Since the former are likely to be absorbed AGN, this may provide support to unification theories, in which obscuration is due to an orientation effect and has nothing to do with the large-scale clustering whatsoever. Other works (e.g. Gilli et al. \cite{Gilli05}, \cite{Gilli09}) have also failed to find significant differentes in the spatial clustering of absorbed and unabsorbed AGN.

We inverted Limber's equation, assuming a given redshift distribution for our sources, in order to estimate typical spatial correlation lengths. We found values of $r_0=$$12.25\pm0.12$, $9.9\pm2.4$, and $7.0\pm5.5$ $h^{-1}$~Mpc in the soft, hard and ultrahard bands, respectively, for a clustering model constant in comoving coordinates, while for clustering constant in physical coordinates we obtained $r_0=$$6.54\pm0.06$, $5.7\pm1.4$, and $5.1\pm4.1$ $h^{-1}$~Mpc, respectively.

Inverting Limber's equation for different flux-limited subsamples reveals no dependence of the typical deprojected spatial length neither on the median luminosity nor on the median redshift of the samples. A slightly positive trend might be observed when assuming a $\epsilon=-3$ clustering, although it does not provide a significantly better fit compared with a constant model, according to the F-test. These results appear to be in contradiction with those of Plionis et al. (\cite{Plionis08}) but are in agreement with those of Gilli et al. (\cite{Gilli05}) and (\cite{Gilli09}). Moreover, this could mean that the $\theta_0$-flux limit dependency discussed in Section~\ref{fluxlimit} might be caused by the fact that pairs of sources tend to appear closer in deep surveys with fainter flux limits.

We used these values to calculate the $rms$ fluctuations of the AGN distributions within a sphere of radius 8 $h^{-1}$~Mpc, and compared them with that of the underlying mass distribution from the linear theory in order to estimate the bias parameter of our X-ray sources. We obtained values ranging from $\sim$2 to $\sim$4.8 in the redshift interval 0.5$\lesssim z \lesssim$1. The bias depends on the mass of the dark matter haloes (DMH) that host the AGN population. From the computed bias values we have estimated a typical DMH mass of $\langle \log M_{DMH}\rangle \simeq 12.60\pm0.34$~$h^{-1}$~M$_\odot$.

The typical AGN lifetime derived from the Press-Schechter approximation at redshift $z \sim 1$ lies in the range in the range $t_{AGN}=3.1-4.5 \times 10^8$~yr. This interval is significantly shorter than the time span between that redshift and the present thus suggesting the existence of many AGN generations, and that a significant fraction of galaxies may switch from a quiescent phase to AGN activity, and vice versa, several times throughout their lifes.

%_________________________________________________

\begin{acknowledgements}
The author acknowledges funding from the European Commission under the Marie Curie Host Fellowships Action for Early Stage Research Training SPARTAN programme (Centre of Excellence for Space, Planetary and Astrophysics Research Training and Networking) Contract No MEST-CT-2004-007512, University of Leicester, UK. FJC acknowledges financial support for this work provided by the Spanish Ministerio de Educaci\'on y Ciencia under project ESP2006-13608-C02-01. SM acknowledges direct support from the UK STFC Research Council. We thank the referee Dr. Spyros Basilakos for useful comments that significantly improved the scientific impact of this paper.
\end{acknowledgements}

%----------------------------------------------------------

\end{document}